\documentclass{article} %
\PassOptionsToPackage{numbers,sort&compress}{natbib}

\usepackage[final]{nips_2017}

\usepackage{import, subfiles}
\usepackage{url}
\usepackage{xr}
\usepackage{tikz, pgf}
\usetikzlibrary{calc}
\usetikzlibrary{decorations.pathreplacing}
\usepackage{amsmath, amsthm, bm, bbm, amssymb}
\usepackage{mathtools}
\usepackage[inline]{enumitem}

\usepackage{amsfonts}

\usepackage{algpseudocode, algorithm}
\algtext*{EndWhile}%
\algtext*{EndIf}%
\algtext*{EndFor}%
\usepackage{multirow}
\usepackage{wrapfig}
\usepackage{graphicx} %
\usepackage{natbib}
\usepackage[colorlinks,citecolor=blue,urlcolor=black,linkcolor=blue,pdfborder={0 0 0}]{hyperref}
\usepackage{todonotes}

\usepackage[capitalize,sort]{cleveref}  

\crefname{nlem}{Lemma}{Lemmas}
\crefname{nprop}{Proposition}{Propositions}
\crefname{ncor}{Corollary}{Corollaries}
\crefname{nthm}{Theorem}{Theorems}
\crefname{assumption}{Assumption}{Assumptions}
\crefname{exa}{Example}{Examples}

\usepackage{color}

\usepackage[arxiv]{optional}

\usepackage{jhh-misc2}
\boldshortcuts

\opt{nips}{\externaldocument{supplementary-material-nips}}

\algrenewcommand{\algorithmiccomment}[1]{\hfill$\triangleright$ #1}

\usepackage{caption}
\usepackage{subcaption}

\graphicspath{ {figures/} }
\allowdisplaybreaks[3]

\makeatletter
\newcommand{\inlineitem}[1][]{%
\ifnum\enit@type=\tw@
    {\descriptionlabel{#1}}
  \hspace{\labelsep}%
\else
  \ifnum\enit@type=\z@
       \refstepcounter{\@listctr}\fi
    \quad\@itemlabel\hspace{\labelsep}%
\fi}
\makeatother
\newcommand{\ii}{\textrm{i}}

\newcommand{\pass}{PASS-GLM\xspace}
\newcommand{\passlr}{PASS-LR2\xspace}

\newcommand{\data}{\mcD}
\newcommand{\param}{\boldsymbol\theta}
\newcommand{\parami}[1][i]{\theta_{#1}}
\newcommand{\meanparami}[1][i]{\bar\theta_{#1}}
\newcommand{\xn}[1][n]{\bx_{#1}}
\newcommand{\xall}{\bX}
\newcommand{\yn}[1][n]{y_{#1}}
\newcommand{\yall}{\by}

\newcommand{\zn}[1][n]{z_{#1}}

\newcommand{\suffstat}{\bt}

\newcommand{\reparam}{\boldsymbol\eta}

\newcommand{\prior}{\pi_{0}}
\newcommand{\post}{\pi_{\data}}
\newcommand{\logpost}{\varpi}
\newcommand{\apost}{\tpi_{\data}}
\newcommand{\llik}{\mcL_{\data}}
\newcommand{\allik}{\tilde\mcL_{\data}}
\newcommand{\efllik}{\mcL_{\data,\text{EF}}}

\newcommand{\link}{g}
\newcommand{\invlink}{\link^{-1}}
\newcommand{\glmMap}{\phi}
\newcommand{\glmMapM}[1][M]{\glmMap_{#1}}
\newcommand{\namedGLMmap}[1]{\glmMap_{\mathsf{#1}}}
\newcommand{\logitMap}{\namedGLMmap{logit}}

\newcommand{\poissonMap}{\namedGLMmap{Poisson}}
\newcommand{\laplaceMap}{\namedGLMmap{Laplace}}
\newcommand{\cauchyMap}{\namedGLMmap{Cauchy}}
\newcommand{\huberMap}{\namedGLMmap{Huber}}
\newcommand{\smHuberMap}{\namedGLMmap{SHuber}}
\newcommand{\basemeasure}{\varsigma}
\newcommand{\basis}[1][m]{\psi_{#1}}

\newcommand{\polyMapCoeffsM}[1][m]{b^{(M)}_{#1}}
\newcommand{\polyCoeffs}[1][m]{b_{#1}}

\newcommand{\paramspace}{\Theta}
\newcommand{\covspace}{\mcX}
\newcommand{\obsspace}{\mcY}

\newcommand{\map}{\param_{\text{MAP}}}
\newcommand{\amap}{\tilde\param_{\text{MAP}}}
\newcommand{\mean}{\bar\param}
\newcommand{\acov}{\tilde\bSigma}

\newcommand{\ball}{\mathbb{B}}

\newcommand{\dw}{d_{\mcW}}
\newcommand{\err}{\text{err}}

\newcommand{\dsname}[1]{\textsc{#1}\xspace}
\newcommand{\webspam}{\dsname{Webspam}}
\newcommand{\covtype}{\dsname{CovType}}
\newcommand{\chemre}{\dsname{ChemReact}}
\newcommand{\codrna}{\dsname{CodRNA}}
\newcommand{\criteo}{\dsname{Criteo}}

\usepackage{autonum}

\makeatletter
\newcommand\footnoteref[1]{\protected@xdef\@thefnmark{\ref{#1}}\@footnotemark}
\makeatother

\title{PASS-GLM: polynomial approximate sufficient statistics for scalable Bayesian GLM inference}
\author{
Jonathan H.~Huggins \\
CSAIL, MIT \\
\texttt{jhuggins@mit.edu}
\And 
Ryan P.~Adams  \\
Google Brain and Princeton  \\
\texttt{rpa@princeton.edu}
\And 
Tamara Broderick \\
CSAIL, MIT \\
\texttt{tbroderick@csail.mit.edu}
}

\begin{document}

\maketitle

\begin{abstract}

Generalized linear models (GLMs)---such as logistic regression, Poisson regression, and robust regression---provide interpretable models for diverse data types. Probabilistic approaches, particularly Bayesian ones, allow coherent estimates of uncertainty, incorporation of prior information, and sharing of power across experiments via hierarchical models. In practice, however, the approximate Bayesian methods necessary for inference have either failed to scale to large data sets or failed to provide theoretical guarantees on the quality of inference. We propose a new approach based on constructing polynomial approximate sufficient statistics for GLMs (\pass). We demonstrate that our method admits a simple algorithm as well as trivial streaming and distributed extensions that do not compound error across computations. We provide theoretical guarantees on the quality of point (MAP) estimates, the approximate posterior, and posterior mean and uncertainty estimates. We validate our approach empirically in the case of logistic regression using a quadratic approximation and show competitive performance with stochastic gradient descent, MCMC, and the Laplace approximation in terms of  speed and multiple measures of accuracy---including on an advertising data set with 40 million data points and 20,000 covariates.
\end{abstract}

\section{Introduction} \label{sec:intro}

Scientists, engineers, and companies increasingly use large-scale data---often only available via streaming---to obtain insights into
their respective problems. For instance, scientists might be interested in understanding how varying experimental inputs leads
to different experimental outputs; or medical professionals might be interested in understanding which elements of patient histories 
lead to certain health outcomes. 
\emph{Generalized linear models} (GLMs) enable these practitioners to explicitly
and interpretably model the effect of
covariates on outcomes while allowing flexible noise distributions---including binary, count-based, and 
heavy-tailed observations. Bayesian approaches further facilitate (1) understanding the importance of covariates via
coherent estimates of parameter uncertainty, (2) incorporating prior knowledge into the analysis, and (3) sharing of power across different
experiments or domains via hierarchical modeling. In practice, however, an exact Bayesian analysis is computationally infeasible
for GLMs, so an approximation is necessary. While some approximate methods provide asymptotic guarantees on quality,
these methods often only run successfully in the small-scale data regime. In order to run on (at least) millions of data points and thousands of covariates,
practitioners often turn to heuristics with no theoretical guarantees on quality. 
In this work, we propose a novel and simple approximation framework for probabilistic inference in GLMs.
We demonstrate theoretical guarantees on the quality of point estimates in the finite-sample setting
and on the quality of Bayesian posterior approximations
produced by our framework. We show that our framework trivially extends to streaming data and to distributed architectures,
with \emph{no} additional compounding of error in these settings. We empirically demonstrate the practicality of
our framework on datasets with up to tens of millions of data points and tens of thousands of covariates.

\textbf{Large-scale Bayesian inference.} Calculating accurate approximate Bayesian posteriors for large data sets together with complex models and potentially high-dimensional
parameter spaces is a long-standing problem. 
We seek a method that satisfies the following criteria: 
(1)  it provides a \emph{posterior approximation};
(2)  it is \emph{scalable}; 
(3)  it comes equipped with \emph{theoretical guarantees}; and
(4)  it provides \emph{arbitrarily good approximations}.
By \emph{posterior approximation} we mean that the method outputs an approximate posterior
distribution, not just a point estimate.
By \emph{scalable} we mean that the method examines each data point only a small number
of times, and further can be applied to streaming and distributed data.
By \emph{theoretical guarantees} we mean that the posterior approximation 
is certified to be close to the true posterior in terms of, for example, some metric on probability measures.  
Moreover, the distance between the exact and approximate posteriors is an efficiently 
computable quantity.
By an \emph{arbitrarily good approximation} we mean that, with a large enough 
computational budget, the method can output an approximation that is as close to the
exact posterior as we wish. 

Markov chain Monte Carlo (MCMC) methods provide an approximate posterior,
and the approximation typically becomes arbitrarily good as the amount of computation time grows asymptotically; thereby 
MCMC satisfies criteria 1, 3, and 4. But scalability of MCMC can be an issue.
Conversely, variational Bayes (VB) and expectation propagation (EP) \citep{Minka:2001we} have grown in popularity
due to their scalability to large data and models---though they typically lack guarantees on quality (criteria 3 and 4). 
Subsampling methods have been proposed to speed up
MCMC~\citep{Welling:2011,Ahn:2012,Bardenet:2014,Korattikara:2014,Maclaurin:2014,Bardenet:2015} and
VB~\citep{Hoffman:2013}. 
Only a few of these algorithms preserve guarantees asymptotic in time (criterion 4), and they often require restrictive assumptions. %
On the scalability front (criterion 2), many though not all subsampling MCMC methods have been found to require examining a constant fraction of the data
at each iteration~\citep{Pillai:2014,Bardenet:2015,Teh:2016,Betancourt:2015,Alquier:2016,Pollock:2016},
so the computational gains are limited.
Moreover, the random data access required by these methods may be
infeasible for very large datasets that do not fit into memory. Finally, they do not apply
to streaming and distributed data, and thus fail criterion 2 above. 
More recently, authors have proposed subsampling methods based on piecewise deterministic Markov processes (PDMPs)~\citep{Bierkens:2016a,BouchardCote:2016,Pakman:2016}. These methods are promising since
subsampling data here does not change the invariant distribution of the continuous-time Markov
process.
 But these methods have not yet been validated on large datasets nor is it understood how subsampling affects the mixing rates of the Markov processes.
Authors have also proposed methods for coalescing information across distributed computation (criterion 2) in
MCMC~\citep{Scott:2013,Srivastava:2015,Rabinovich:2015,Entezari:2016}, 
VB~\citep{Broderick:2013b,Campbell:2015}, and EP~\citep{Gelman:2014,Teh:2015}---and in the case of VB, across epochs
as streaming data is collected~\citep{Broderick:2013b,Campbell:2015}.
(See \citet{Angelino:2016} for a broader discussion of issues surrounding scalable Bayesian inference.)
While these methods lead to gains in computational efficiency, they lack rigorous justification 
and provide no guarantees on the quality of inference (criteria 3 and 4).

To address these difficulties, we are inspired in part by the observation that not all Bayesian models require
expensive posterior approximation. When the likelihood belongs to an \emph{exponential family}, Bayesian posterior
computation is fast and easy. In particular, it suffices to find the \emph{sufficient statistics} of the data, which
require computing a simple summary at each data point and adding these summaries across data points. The latter addition
requires a single pass through the data and is trivially streaming or distributed. With the sufficient statistics in hand, the posterior
can then be calculated via, e.g., MCMC,
and point estimates such as the MLE can be computed---all in time independent of the data set size.
Unfortunately, sufficient statistics are not generally available (except in very special cases) for GLMs. 
We propose to instead develop a notion of \emph{approximate sufficient statistics}. 
Previously authors have suggested using a \emph{coreset}---a weighted data subset---as a summary of the
data~\citep{Feldman:2011b,Fithian:2014,Lucic:2017,Huggins:2016,Han:2016,Bachem:2017}.
While these methods provide theoretical guarantees on the quality of inference
via the model evidence, the 
resulting guarantees are better suited to approximate optimization and do not translate to guarantees on typical
Bayesian desiderata, such as the accuracy of posterior mean and uncertainty estimates.
Moreover, while these methods do admit streaming and distributed constructions, the approximation error
is compounded across computations. 

\textbf{Our contributions.} In the present work we instead propose to construct our approximate sufficient statistics via a much simpler
\emph{polynomial approximation}
for generalized linear models. 
We therefore call our method 
\emph{polynomial approximate sufficient statistics for generalized linear models} (\pass).
\pass satisfies all of the criteria laid of above.
It provides a \emph{posterior approximation} with \emph{theoretical guarantees} (criteria 1 and 3).
It is \emph{scalable} since is requires only a single pass over the data and can be 
applied to streaming and distributed data (criterion 2).
And by increasing the number of approximate sufficient statistics, \pass can produce
\emph{arbitrarily good approximations} to the posterior (criterion 4).

The Laplace approximation \citep{tierney1986accurate} and variational methods with a
Gaussian approximation family \citep{jaakkola1997variational,Kucukelbir:2015vs}
may be seen as polynomial (quadratic) approximations in the log-likelihood space.
But we note that the VB variants still suffer the issues described above.
A Laplace approximation relies on a Taylor series expansion of the log-likelihood around
the \emph{maximum a posteriori} (MAP) solution, 
which requires first calculating the MAP---an expensive multi-pass optimization 
in the large-scale data setting. Neither Laplace nor VB offers the simplicity of sufficient statistics, including in streaming and
distributed computations.
The recent work of \citet{Stephanou:2017} is similar in spirit to ours, though they address a different statistical problem: they construct sequential quantile estimates using Hermite polynomials.

In the remainder of the paper, we begin by describing generalized linear models in more detail in \cref{sec:background}. We construct
our novel polynomial approximation and specify our \pass algorithm in \cref{sec:pass}. 
We will see that streaming and distributed computation are trivial for our algorithm and do
not compound error.
In \cref{sec:map-theory}, we demonstrate finite-sample guarantees on the quality of the
MAP estimate arising from our algorithm, with the maximum likelihood
estimate (MLE) as a special case.
In \cref{sec:bayes-theory}, we prove guarantees on the Wasserstein distance between the exact and approximate posteriors---and
thereby bound both posterior-derived point estimates and uncertainty estimates.
In \cref{sec:experiments}, we demonstrate the efficacy of our approach in practice by focusing on logistic regression.
We demonstrate experimentally that \pass can be scaled with almost no loss of efficiency to multi-core architectures.
We show on a number of real-world datasets---including a large, high-dimensional advertising dataset
(40 million examples with 20,000 dimensions)---that \pass provides an attractive trade-off between computation and 
accuracy.

\section{Background}
\label{sec:background}

\textbf{Generalized linear models.}
\emph{Generalized linear models} (GLMs) combine the interpretability of linear models with the flexibility of more general outcome
distributions---including binary, ordinal, and heavy-tailed observations.
Formally, we let~${\obsspace \subseteq \reals}$ be the observation space,~${\covspace \subseteq \reals^{d}}$ be
the covariate space, and~${\paramspace \subseteq \reals^{d}}$ be the parameter space.
Let~${\data \defined \theset{(\xn, \yn)}_{n=1}^{N}}$ be the observed data.
We write~${\xall \in \reals^{N \times d}}$ for the matrix of all covariates and~${\yall \in \reals^{N}}$ 
for the vector of all observations. 
We consider GLMs
\[
\textstyle\log p(\yall \given \xall, \param) = \sum_{n=1}^{N} \log p(\yn \given \invlink(\xn \cdot \param)) = \sum_{n=1}^{N} \glmMap(\yn, \xn \cdot \param),
\]
where ${\mu \defined \invlink(\xn \cdot \param)}$ is the expected value of~$\yn$ and~${\invlink : \reals \to \reals}$ is the \emph{inverse link function}. %
We call~${\glmMap(\yn[], s) \defined \log p(\yn[] \given \invlink(s))}$ the GLM \emph{mapping function}. 

Examples include some of the most widely used 
models in the statistical toolbox.
For instance, for binary observations ${\yn[] \in \{\pm 1\}}$, the likelihood model is Bernoulli, ${p(\yn[] = 1 \given \mu) = \mu}$, and
the link function is often either the logit~${\link(\mu) = \log \frac{\mu}{1 - \mu}}$ (as in logistic regression) or 
the probit~${\link(\mu) = \Phi^{-1}(\mu)}$, where~$\Phi$ is the standard Gaussian CDF. 
When modeling count data~${\yn[] \in \nats}$, the likelihood model might be Poisson,~${p(\yn[] \given \mu) = \mu^{\yn[]}e^{-\mu}/\yn[]!}$, 
and ${\link(\mu) = \log(\mu)}$ is the typical log link. 
Other GLMs include gamma regression, robust regression, and binomial regression, all of which are commonly used for large-scale data analysis 
(see~\cref{ex:robust-regression,ex:gamma-regression}).

If we place a prior~$\prior(\dee \param)$ on the parameters, then a full Bayesian analysis aims to approximate 
the (typically intractable) GLM posterior distribution $\post(\dee\param)$, where
\[
\post(\dee\param) = \frac{p(\yall \given \xall, \param)\,\prior(\dee\param)}{\int p(\yall \given \xall, \param')\,\prior(\dee\param')}.
\]
The \emph{maximum a posteriori} (MAP) solution gives a point estimate of the parameter:
\[
\map \defined \argmax_{\param \in \paramspace} \post(\param) = \argmax_{\param \in \paramspace} \log\prior(\param) + \llik(\param), \label{eq:map}
\]
where~${\llik(\param) \defined \log p(\yall \given \xall, \param)}$ is the data log-likelihood. 
The MAP problem strictly generalizes finding the maximum likelihood estimate (MLE), since the MAP solution equals the MLE when 
using the (possibly improper) prior~${\prior(\param) = 1}$. 

\textbf{Computation and exponential families.}
In large part due to the high-dimensional integral implicit in the normalizing constant, approximating the posterior, e.g.,
via MCMC or VB, is often prohibitively expensive. Approximating this integral will typically
require many evaluations of the (log-)likelihood, or its gradient, and each evaluation 
may require~$\Omega(N)$ time.

Computation is much more efficient, though, if the model is in an \emph{exponential family} (EF). In the EF case, there 
exist functions~${\suffstat, \reparam : \reals^{d} \to \reals^{m}}$, %
such that\footnote{Our presentation is slightly different from the standard textbook account because we have 
implicitly absorbed the base measure and log-partition function into $\suffstat$ and $\reparam$.}
\[
\log p(\yn \given \xn, \param) 
&= \suffstat(\yn, \xn) \cdot \reparam(\param)
=: \efllik(\param; \suffstat(\yn, \xn)). 
\]
Thus, we can rewrite the log-likelihood as 
\[
\textstyle\llik(\param) = \sum_{n=1}^{N}\efllik(\param; \suffstat(\yn, \xn)) =: \efllik(\param; \suffstat(\data)), 
\]
where~${\suffstat(\data) \defined \sum_{n=1}^{N} \suffstat(\yn, \xn)}$. 
The \emph{sufficient statistics} $\suffstat(\data)$ can be calculated in~$O(N)$ time, after which each 
evaluation of~$\efllik(\param; \suffstat(\data))$ or~$\grad\efllik(\param; \suffstat(\data))$ requires only~$O(1)$ time. Thus, instead of~$K$ passes over~$N$ data (requiring~$O(NK)$ time), only $O(N + K)$ time is needed. 
Even for moderate values of $N$, the time savings can be substantial when $K$ is large.

The Poisson distribution is an illustrative example of a one-parameter exponential family with~${\suffstat(y) = (1, y, \log y!)}$ and~${\reparam(\theta) = (\theta, \log \theta, 1)}$.
Thus, if we have data~$\yall$ (there are no covariates),~${\suffstat(\yall) = (N, \sum_{n}\yn, \sum \log \yn!)}$. 
In this case it is easy to calculate that the maximum likelihood estimate of~$\theta$ from~$\suffstat(\yall)$ as~${t_{1}(\yall) / t_{0}(\yall) = N^{-1}\sum_{n}\yn}$. 

Unfortunately, GLMs rarely belong to an exponential family -- even if the outcome distribution is in an exponential family, the 
use of a link destroys the EF structure. 
In logistic regression, we write (overloading the~$\glmMap$ notation) ${\log p(\yn \given \xn, \param) = \logitMap(\yn\xn \cdot \param)}$, where ${\logitMap(s) \defined -\log(1 + e^{-s})}$. 
For Poisson regression with log link, ${\log p(\yn \given \xn, \param) = \poissonMap(\yn,  \xn \cdot \param)}$, where~${\poissonMap(\yn[], s) \defined \yn[] s - e^{s} - \log \yn[]!}$. 
In both cases, 
we cannot express the log-likelihood as an inner product between a
function solely of the data and a function solely of the parameter.  %

\section{\pass}
\label{sec:pass}

\begin{algorithm}[t]
\caption{\pass inference}\label{alg:pass-glm}
\begin{algorithmic}[1] 
\Require data $\data$, GLM mapping function $\glmMap : \reals \to \reals$, degree $M$, polynomial basis $(\basis)_{m \in \nats}$ 
with base measure $\basemeasure$
\State Calculate basis coefficients $\polyCoeffs \gets \int\glmMap\basis\dee\basemeasure$ using numerical integration for $m=0,\dots,M$
\State Calculate polynomial coefficients $\polyMapCoeffsM \gets \sum_{k=m}^{M}\alpha_{k,m}\polyCoeffs$ for $m=0,\dots,M$
\For{$\bk \in \nats^{d}$ with $\sum_{j}k_{j} \le M$}
	\State Initialize $t_{\bk} \gets 0$
\EndFor
\For{$n=1,\dots,N$} \Comment{Can be done with any combination of batch, parallel, or streaming}
	\For{$\bk \in \nats^{d}$ with $\sum_{j}k_{j} \le M$}
		\State Update $t_{\bk} \gets t_{\bk} + (\yn\xn)^{\bk}$
	\EndFor
\EndFor
\State Form approximate log-likelihood $\allik(\param) =  \sum_{\bk \in \nats^{d} : \sum_{j} k_{j} \le m} { m\choose \bk } \polyMapCoeffsM  t_{\bk}\param^{\bk}$
\State Use $\allik(\param)$ to construct approximate posterior $\apost(\param)$
\end{algorithmic}
\end{algorithm}

Since exact sufficient statistics are not available for GLMs, we propose to construct \emph{approximate sufficient statistics}.
In particular, we propose to approximate the mapping function~$\glmMap$ with an order-$M$ polynomial $\glmMapM$.
We therefore call our method \emph{polynomial approximate sufficient statistics for GLMs} (\pass).
We illustrate our method next in the logistic regression case, where~${\log p(\yn \given \xn, \param) = \logitMap(\yn\xn \cdot \param)}$. 
The fully general treatment appears in \cref{app:pass-glm-general}. 
Let~${\polyMapCoeffsM[0], \polyMapCoeffsM[1] \dots, \polyMapCoeffsM[M]}$ be constants such that 
\[
\textstyle\logitMap(s) \approx \glmMapM(s) \defined \sum_{m=0}^{M}  \polyMapCoeffsM s^{m}.
\]
Let ${\bv^{\bk} \defined \prod_{j=1}^{d} v_{j}^{k_{j}}}$ for vectors~${\bv, \bk \in \reals^{d}}$. 
Taking $s = \yn[]\xn[] \cdot \param$, we obtain
\[
\logitMap(\yn[]\xn[] \cdot \param) 
&\approx \glmMapM(\yn[]\xn[] \cdot \param) 
= \textstyle\sum_{m=0}^{M}  \polyMapCoeffsM(\yn[]\xn[] \cdot \param)^{m} 
= \textstyle\sum_{m=0}^{M} \polyMapCoeffsM \sum_{\substack{\bk \in \nats^{d} \\ \sum_{j} k_{j} = m}} {m \choose \bk } (\yn[]\xn[])^{\bk}\param^{\bk} \\
&= \textstyle\sum_{m=0}^{M} \sum_{\bk \in \nats^{d} : \sum_{j} k_{j} = m} a(\bk, m, M)  (\yn[]\xn[])^{\bk}\param^{\bk},
\]
where ${ m \choose \bk }$ is the multinomial coefficient and ${a(\bk, m, M)  \defined  { m\choose \bk } \polyMapCoeffsM}$. 
Thus, $\glmMapM$ is an $M$-degree polynomial approximation to~$\logitMap(\yn[]\xn[] \cdot \param)$ with the~${d + M \choose d}$
monomials of degree at most~$M$ serving as sufficient statistics derived from~$\yn[]\xn[]$. 
Specifically, we have a exponential family model with 
\[
\suffstat(\yn[]\xn[]) &= ([\yn[]\xn[]]^{\bk})_{\bk} &
\text{and}&&
\reparam(\param) &= (a(\bk, m, M)\param^{\bk})_{\bk},
\]
where $\bk$ is taken over all $\bk \in \nats^{d}$ such that $\sum_{j} k_{j} \le M$. 
We next discuss the calculation of the $\polyMapCoeffsM$ and the choice of $M$.

\textbf{Choosing the polynomial approximation.}
To calculate the coefficients~$\polyMapCoeffsM$, 
we choose a polynomial basis~$(\basis)_{m \in \nats}$ orthogonal with respect to 
a base measure $\basemeasure$, where $\basis$ is degree~$m$~\citep{Szego:1975}. 
That is,~${\basis(s) = \sum_{j=0}^{m}\alpha_{m,j}s^{j}}$ for some~$\alpha_{m,j}$, and~${\int \basis\basis[m']\dee\basemeasure = \delta_{mm'}}$, 
where~${\delta_{mm'} = 1}$ if~${m=m'}$ and zero otherwise.
If~${\polyCoeffs \defined \int \glmMap\basis\dee\basemeasure}$, %
then~${\glmMap(s) = \sum_{m=0}^{\infty} \polyCoeffs \psi_{m}(s)}$
and the approximation~${\glmMapM(s) = \sum_{m=0}^{M} \polyCoeffs \psi_{m}(s)}$. 
Conclude that~${\polyMapCoeffsM = \sum_{k=m}^{M}\alpha_{k,m}\polyCoeffs}$. 
The complete \pass framework appears in \cref{alg:pass-glm}.

Choices for the orthogonal polynomial basis include Chebyshev, Hermite, Leguerre, and Legendre polynomials~\citep{Szego:1975}. 
We choose Chebyshev polynomials since they %
provide a uniform quality guarantee on a finite interval, e.g.,~$[-R, R]$ for some~${R > 0}$ in what follows. 
If $\glmMap$ is smooth, the choice of Chebyshev polynomials (scaled appropriately, along with the base measure~$\basemeasure$, based on the choice of~$R$) yields error exponentially small in~$M$:~${\sup_{s \in [-R,R]} |\glmMap(s) - \glmMapM(s)| \le C\rho^{M}}$ for some~${0 < \rho < 1}$ and~${C > 0}$ \citep{Mason:2003}. 
We show in \cref{app:chebyshev} that the error in the approximate derivative $\glmMapM'$ is also exponentially small in $M$: $\sup_{s \in [-R,R]} |\glmMap'(s) - \glmMapM'(s)| \le C'\rho^{M}$, where $C' > C$. 

\textbf{Choosing the polynomial degree.}
For fixed $d$, the number of monomials is $O(M^{d})$ while for fixed $M$ the number of monomials is~$O(d^{M})$.
The number of approximate sufficient statistics can remain manageable when either $M$ or $d$ is small but becomes unwieldy if $M$ and $d$
are both large. Since our experiments (\cref{sec:experiments}) generally have large $d$, we focus on the small $M$ case here.

In our experiments we further focus on the choice of logistic regression as a particularly popular GLM example
with ${p(\yn \given \xn, \param) = \logitMap(\yn\xn \cdot \param)}$, where ${\logitMap(s) \defined -\log(1 + e^{-s})}$.
In general, the smallest and therefore most compelling
choice of $M$ \emph{a priori is 2}, and we demonstrate the reasonableness of this choice empirically in \cref{sec:experiments} 
for a number of large-scale data analyses.
In addition, in the logistic regression case, $M=6$ is the next usable choice beyond $M=2$. 
This is because
$\polyMapCoeffsM[2k+1] = 0$ for all integer~${k \ge 1}$ with~${2k+1 \le M}$. 
So any approximation beyond~${M=2}$ must have~${M \ge 4}$.
Also, $\polyMapCoeffsM[4k] > 0$ for all integers ${k \ge 1}$ with ${4k \le M}$. 
So choosing~${M=4k}$,~${k \ge 1}$, leads to a pathological approximation of $\logitMap$
where the log-likelihood can be made arbitrarily large by taking~${\|\param\|_{2} \to \infty}$.
Thus, a reasonable polynomial approximation for logistic regression requires~${M=2+4k}$,~${k \ge 0}$.
We have discussed the relative drawbacks of other popular quadratic approximations, including the
Laplace approximation and variational methods, in \cref{sec:intro}.

\section{Theoretical Results}

We next establish quality guarantees for \pass. We first provide finite-sample and asymptotic guarantees on the MAP (point estimate) solution, and therefore on the MLE, in \cref{sec:map-theory}. We then provide guarantees on the Wasserstein distance between the approximate and exact posteriors, and show these bounds translate into bounds on the quality of posterior mean and uncertainty estimates, in \cref{sec:bayes-theory}.
See \cref{app:proofs} for extended results, further discussion, and all proofs.

\subsection{MAP approximation}
\label{sec:map-theory}

In \cref{app:proofs}, we state and prove \cref{thm:approx-map}, which provides guarantees on the quality of the MAP estimate for an arbitrary approximation
$\allik(\param)$
to the log-likelihood $\llik(\param)$. The approximate MAP (i.e., the MAP under~$\allik$) is (cf.~\cref{eq:map})
\[ 
\amap \defined \argmax_{\param \in \paramspace}  \log\prior(\param)  + \allik(\param). 
\]
Roughly, we find in \cref{thm:approx-map} that the error in the MAP estimate naturally depends on the error of the approximate log-likelihood
as well as the peakedness of the posterior near the MAP.
In the latter case, if $\log \post$ is very flat, then even a small error from using $\allik$ in
place of $\llik$ could lead to a large error in the approximate MAP solution. 
We measure the peakedness of the distribution in terms of the strong convexity constant\footnote{Recall that a twice-differentiable function $f : \reals^{d} \to \reals$ is $\varrho$-strongly convex 
at $\param$ if the minimum eigenvalue of the Hessian of $f$ evaluated at $\param$ is at least $\varrho > 0$.} 
of $-\log \post$ near $\map$. 

We apply \cref{thm:approx-map} to \pass for logistic regression and robust regression. 
We require the assumption that 
\[
\glmMapM(t) \le \glmMap(t)~\forall t \notin [-R,R], \label{eq:bounding-assumption}
\]
which in the cases of logistic regression and smoothed Huber regression, we conjecture holds for~${M = 2 + 4k}$, ${k \in \nats}$.
For a matrix $\bA$, $\|\bA\|_{2}$ denotes its spectral norm.

\bncor \label{cor:logit-map-error}
For the logistic regression model, assume that ${\|(\grad^{2}\llik(\map))^{-1}\|_{2} \le cd/N}$ for some constant~$c > 0$ 
and that ${\|\xn\|_{2} \le 1}$ for all $n=1,\dots,N$. %
Let $\glmMapM$ be the order-$M$ Chebyshev approximation to $\logitMap$ on~$[-R,R]$ such that \cref{eq:bounding-assumption} holds. %
Let~$\apost(\param)$ denote the posterior approximation obtained by using~$\glmMapM$ with a log-concave prior. 
Then there exist numbers $r = r(R) > 1$, $\veps = \veps(M) = O(r^{-M})$, and 
${\alpha^{*} \ge \frac{27}{\veps d^{3}c^{3} + 54}}$, such that if
${R - \|\map\|_{2} \ge 2\sqrt{\frac{cd \veps}{\alpha^{*}}}}$, then 
\[
\|\map - \amap\|_{2}^{2} 
\le \frac{4cd \veps}{\alpha^{*}} 
\le \frac{4}{27}c^{4}d^{4}\veps^{2} + 8cd\veps. \label{eq:logit-map-error-bound}
\]
\encor

The main takeaways from \cref{cor:logit-map-error} are that 
(1) the error decreases exponentially in $M$ thanks to the $\veps$ term,
(2) the error does not depend on the amount of data, and
(3) in order for the bound on the approximate MAP solution to hold, the norm of the true MAP solution
must be sufficiently smaller than $R$. 

\bnrmk 
Some intuition for the assumption on the Hessian of $\llik$,
i.e., $\textstyle\grad^{2}\llik(\param) = \sum_{n=1}^{N} \logitMap''(\yn\xn \cdot \param) \xn\xn^{\top}$,
is as follows.
Typically for $\param$ near~$\map$, the minimum eigenvalue of $\grad^{2}\llik(\param)$ is 
at least $N/(cd)$ for some ${c > 0}$.
The minimum eigenvalue condition in \cref{cor:logit-map-error} holds if, for example, a constant fraction of the data
satisfies ${0 < b \le \|x_{n}\|_{2} \le  B < \infty}$ and that subset of the data 
does not lie too close to any $(d-1)$-dimensional hyperplane. 
This condition essentially requires the data not to be degenerate
and is similar to ones used to show asymptotic consistency of logistic regression~\citep[Ex.~5.40]{vanderVaart:1998}.
\enrmk

The approximate MAP error bound in the robust regression case using, for example, the smoothed Huber loss~(\cref{ex:robust-regression}), 
is quite similar to the logistic regression result.  

\bncor \label{cor:huber-map-error}
For robust regression with smoothed Huber loss, assume that a constant fraction of the data satisfies $|\xn \cdot \map - \yn| \le b/2$
and that ${\|\xn\|_{2} \le 1}$ for all $n=1,\dots,N$.
Let $\glmMapM$ be the order $M$ Chebyshev approximation to $\huberMap$ on $[-R,R]$ such that \cref{eq:bounding-assumption} holds. %
Let $\apost(\param)$ denote the posterior approximation obtained by using $\glmMapM$ with a log-concave prior. 
Then if $R \gg \|\map\|_{2}$, there exists $r > 1$ such that for $M$ sufficiently large, 
$
\|\map - \amap\|_{2}^{2} = O(d r^{-M}). \label{eq:huber-map-error-bound}
$
\encor

\subsection{Posterior approximation}
\label{sec:bayes-theory}

We next establish guarantees on how close the approximate and exact posteriors are in \emph{Wasserstein distance}, $\dw$. 
For distributions $P$ and~$Q$ on $\reals^{d}$, ${\dw(P, Q) \defined \sup_{f : \|f\|_{L} \le 1}|\int f \dee P - \int f \dee Q|}$,
where $\|f\|_{L}$ denotes the Lipschitz constant of $f$.\footnote{The Lipschitz constant of function ${f :\reals^d \to \reals}$ is ${\|f\|_L \defined  \sup_{\bv, \bw \in \reals^{d}} \frac{\|\phi(\bv) - \phi(\bw)\|_2}{\|\bv-\bw\|_2}}$.}
This choice of distance is particularly useful since, if ${\dw(\post, \apost) \le \delta}$, then $\apost$ can be used to estimate any function with bounded gradient with error at most ${\delta \sup_{\bw}\|\grad f(\bw)\|_{2}}$. 
Wasserstein error bounds therefore give bounds on the mean estimates (corresponding to~${f(\param) = \parami}$) as well as 
uncertainty estimates such as mean absolute deviation (corresponding to~${f(\param) = |\meanparami - \parami|}$, where~$\meanparami$ is the expected value of~$\parami$). 

Our general result (\cref{thm:approx-post}) is stated and proved in \cref{app:proofs}.
Similar to \cref{thm:approx-map}, the result primarily depends on the peakedness of the approximate posterior 
and the error of the approximate gradients. 
If the gradients are poorly approximated then the error can be large 
while if the (approximate) posterior is flat then even small gradient errors could lead to large shifts in expected values of the parameters
and hence large Wasserstein error. 

We apply \cref{thm:approx-post} to \pass for logistic regression and Poisson regression. 
We give simplified versions of these corollaries in the main text and defer the more
detailed versions to \cref{app:proofs}. 
For logistic regression we assume $M=2$ and $\paramspace = \reals^{d}$ since this is the setting we use for our experiments. 
The result is similar in spirit to \cref{cor:logit-map-error}, though more straightforward since $M=2$. 
Critically, we see in this result how having small error depends on $|\yn\xn \cdot \mean|  \le R$ with high probability. 
Otherwise the second term in the bound will be large. 

\bncor \label{cor:approx-post-logistic-simplified}
Let $\glmMapM[2]$ be the second-order Chebyshev approximation to $\logitMap$ on $[-R,R]$ 
and let~${\apost(\param) = \distNorm(\param \given \amap, \acov)}$ denote the posterior approximation obtained by using~$\glmMapM[2]$ 
with a Gaussian prior~${\prior(\param) = \distNorm(\param \given \param_{0}, \bSigma_{0})}$. 
Let~${\mean \defined \int \param \post(\dee\param)}$, let~${\delta_{1} \defined N^{-1}\sum_{n=1}^{N}\ip{\yn\xn}{\mean}}$,
and let~$\sigma_{1}$ be the subgaussianity constant of the random variable~${\ip{\yn\xn}{\mean} - \delta_{1}}$, where~${n \dist \distUnif\theset{1,\dots,N}}$. 
Assume that~${|\delta_{1}| \le R}$, that~${\|\acov\|_{2} \le cd/N}$, and 
that~${\|\xn\|_{2} \le 1}$  for all~${n=1,\dots,N}$. 
Then with~${\sigma_{0}^{2} \defined \|\bSigma_{0}\|_{2}}$, we have 
\[
\dw(\post, \apost) = O\left(dR^{4} + d\sigma_{0} \exp\left({\sigma_{1}^{2}}{\sigma_{0}^{-2}}-\sqrt{2}\sigma_{0}^{-1}(R - |\delta_{1}|)\right)\right)\,.
\] 
\encor

\begin{figure}[tb]
\begin{center}
\begin{subfigure}[b]{.4\textwidth}
    \includegraphics[width=\textwidth]{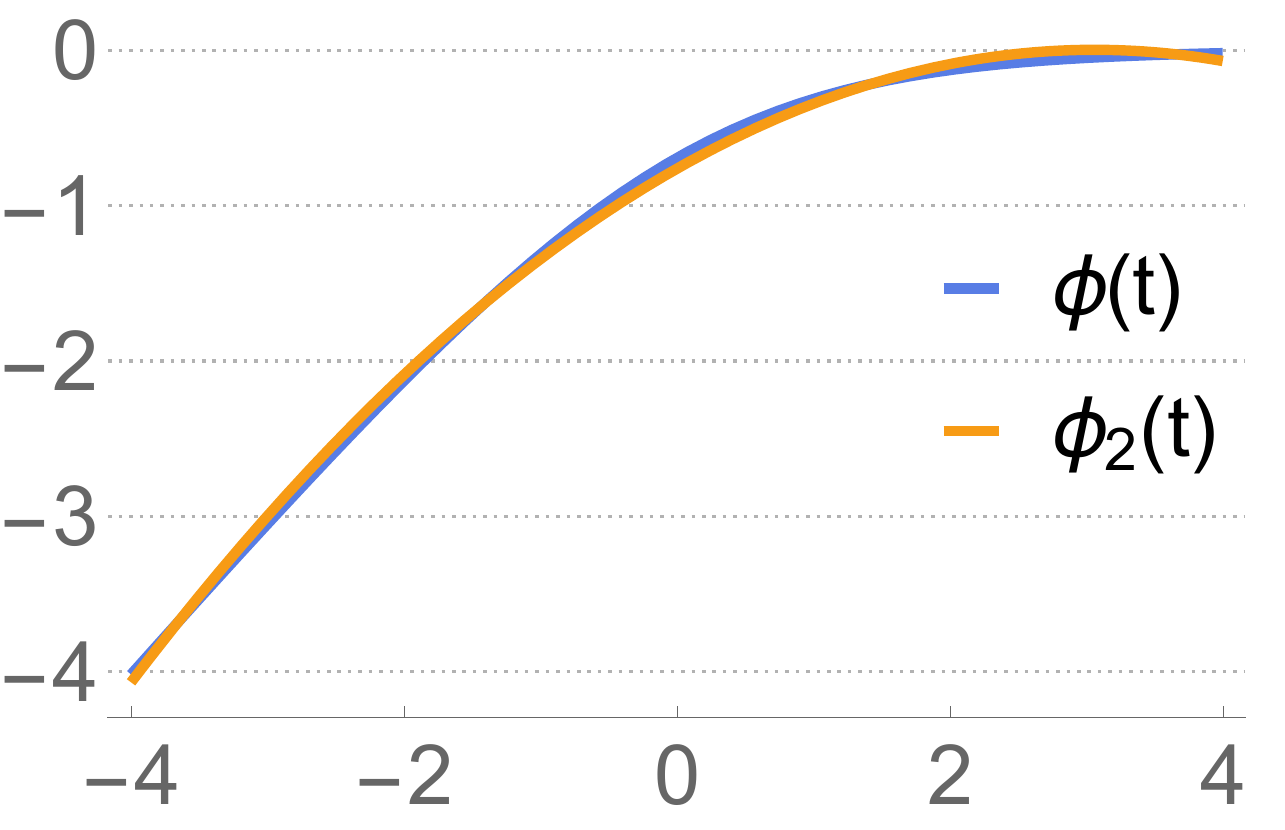} \\
    \caption{}
    \label{fig:2nd-degree-approx}
\end{subfigure}
\begin{subfigure}[b]{.49\textwidth}
    \includegraphics[width=.49\textwidth]{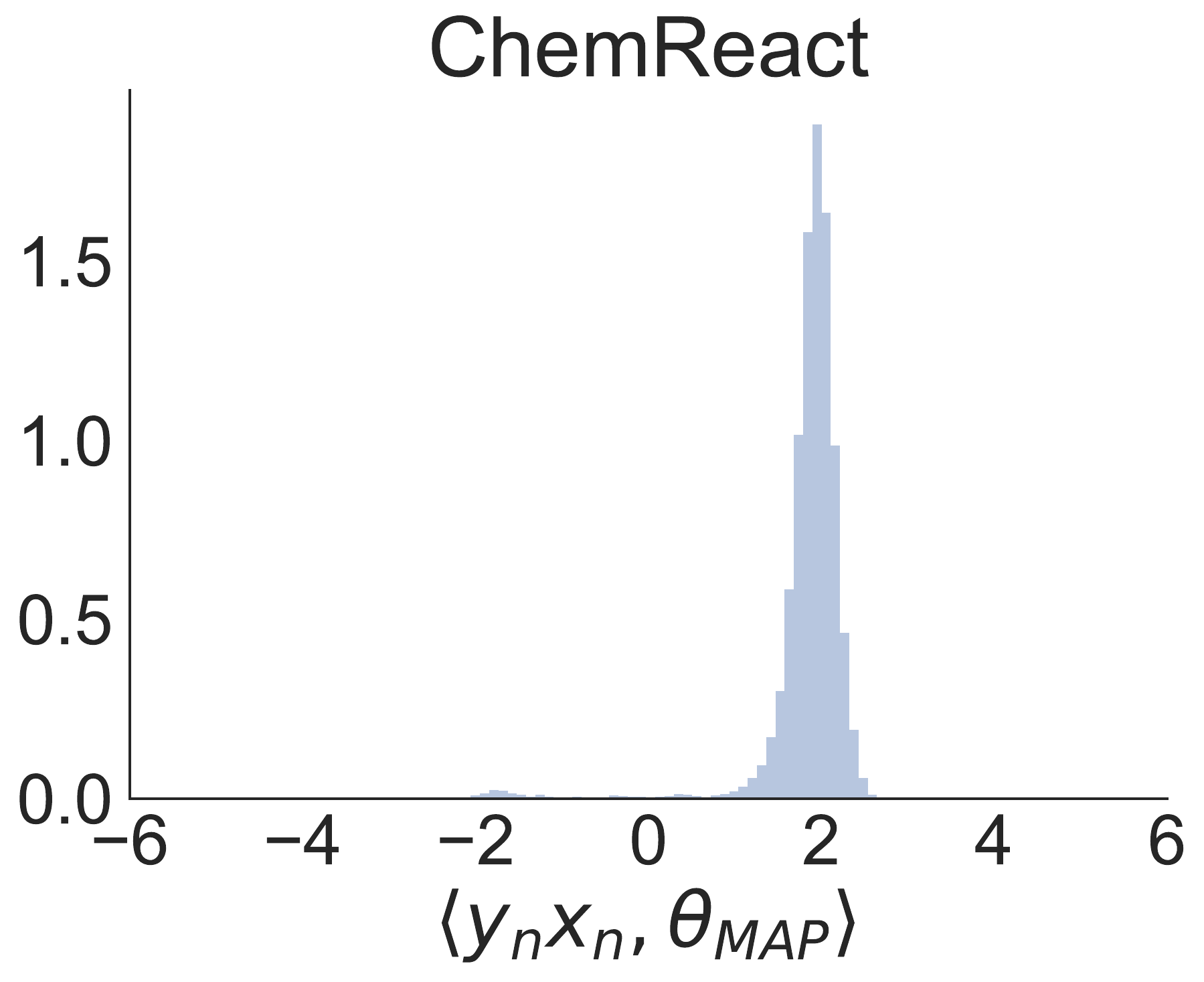} 
    \includegraphics[width=.49\textwidth]{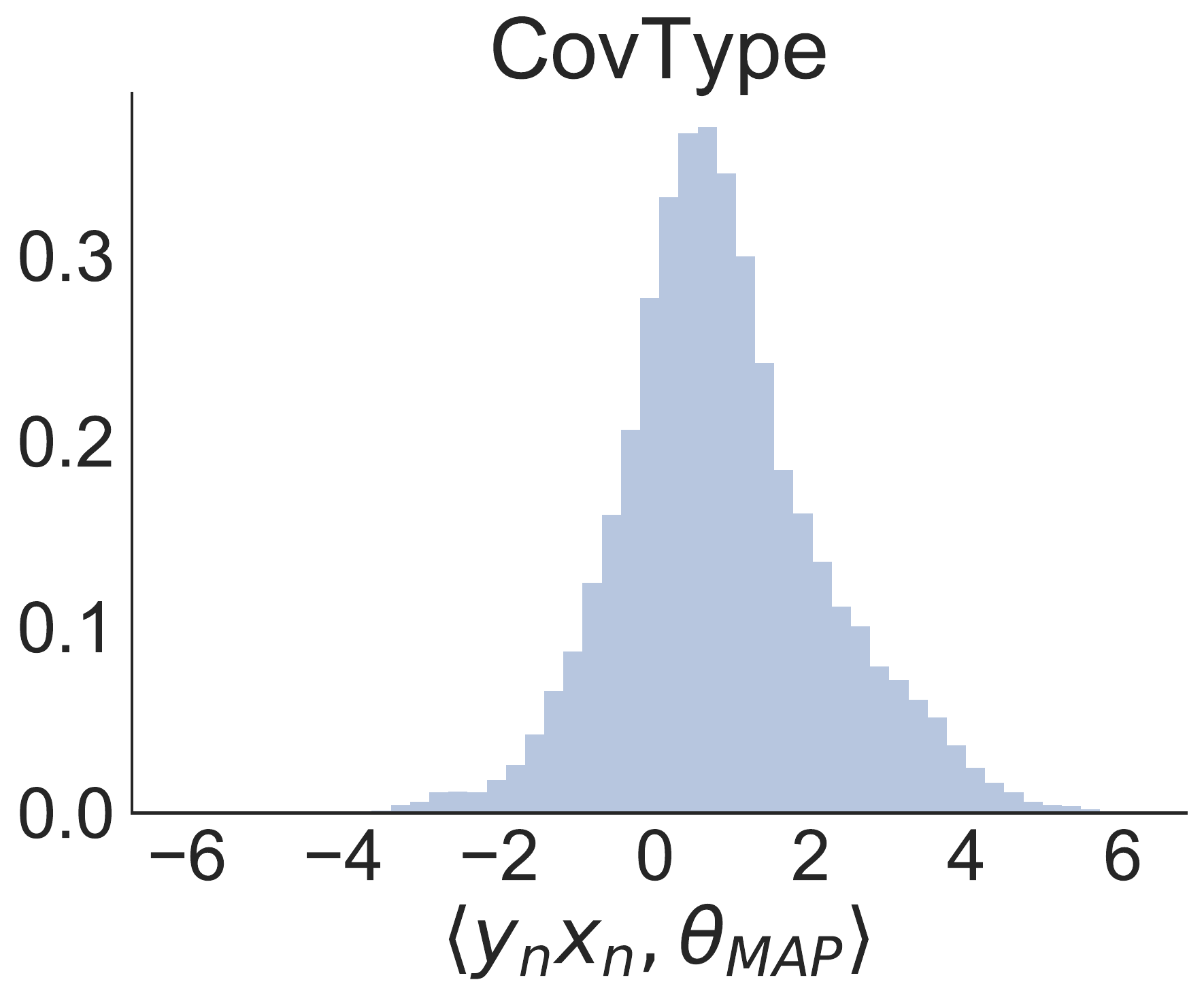}\\
    \includegraphics[width=.49\textwidth]{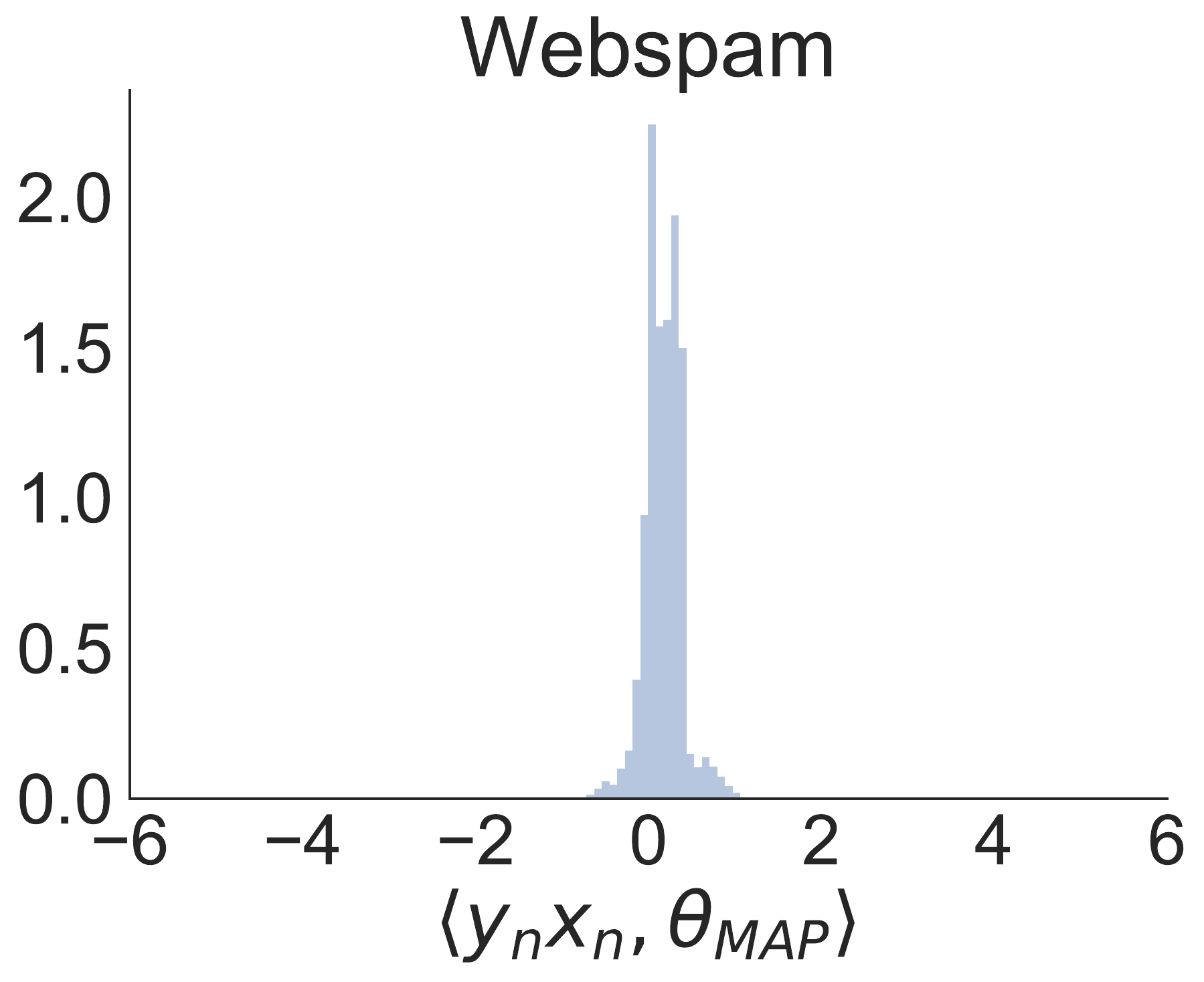}
    \includegraphics[width=.49\textwidth]{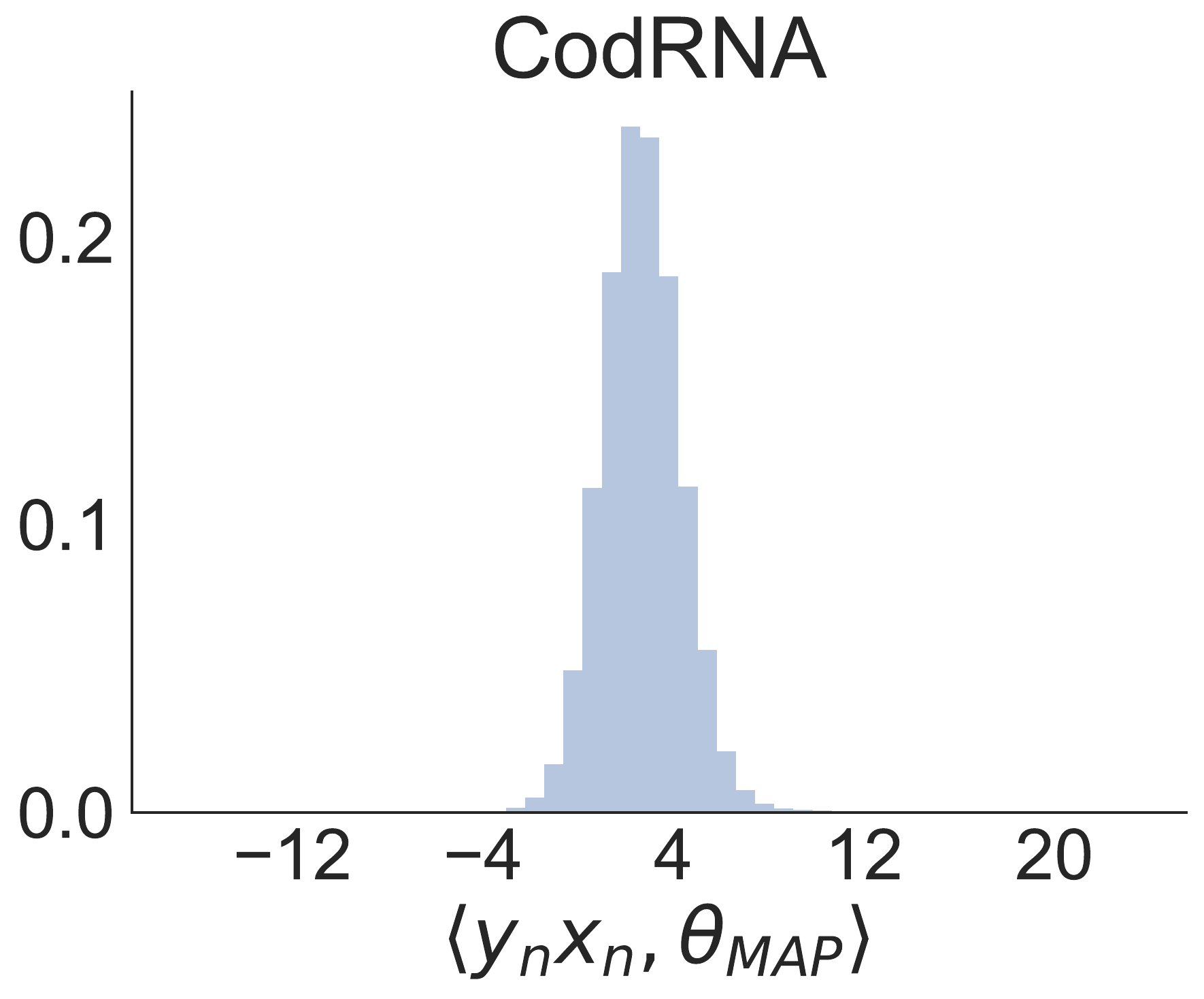}
    \caption{}
    \label{fig:mode-inner-prods}
\end{subfigure}
\vspace{.5em}
\caption{Validating the use of \pass with $M=2$. 
\textbf{(a)} The second-order Chebyshev approximation to $\glmMap = \logitMap$ on $[-4,4]$ is very accurate, with error of at most 0.069. 
\textbf{(b)} For a variety of datasets, the inner products $\ip{\yn\xn}{\map}$ are mostly in the range of $[-4,4]$.}
\label{fig:second-order-stuff}
\end{center}
\end{figure}

The main takeaway from \cref{cor:approx-post-logistic-simplified} is that if 
(a) for most $n$, $|\ip{\xn}{\mean}| < R$, so that $\glmMapM[2]$ is a good approximation to $\logitMap$, and 
(b) the approximate posterior concentrates quickly, then we get a high-quality approximate posterior. 
This result matches up with the experimental results (see \cref{sec:experiments} for further discussion).

For Poisson regression, we return to the case of general $M$.
Recall that in the Poisson regression model that the expectation of $\yn$ is 
${\mu = e^{\xn \cdot \param}}$. 
If $\yn$ is bounded and has non-trivial probability of being greater than zero, we lose little by
restricting $\xn \cdot \param$ to be bounded. 
Thus, we will assume that the parameter space is bounded. 
As in \cref{cor:logit-map-error,cor:huber-map-error}, the error is exponentially small in $M$ and, 
as long as $\|\sum_{n=1}^{N} \xn\xn^{\top}\|_{2}$ grows linearly in $N$, does not depend on the amount of data. 

\bncor \label{cor:approx-post-poisson-simplified}
Let $f_{M}(s)$ be the order-$M$ Chebyshev approximation to $e^{t}$ on the interval~${[-R,R]}$,
and let~$\apost(\param)$ denote the posterior approximation obtained by using the approximation
${\log \tp(\yn \given \xn, \param)  \defined  \yn \xn \cdot \param - f_{M}(\xn \cdot \param) - \log \yn!}$
with a log-concave prior on~${\paramspace = \ball_{R}(\bzero)}$.
If ${\inf_{s \in [-R,R]} f_{M}''(s) \ge \tilde\varrho > 0}$, $\|\sum_{n=1}^{N} \xn\xn^{\top}\|_{2} = \Omega(N/d)$,
and $\|\xn\|_{2} \le 1$  for all $n=1,\dots,N$, then
\[
\dw(\post, \apost) = O\left(d\tilde\varrho^{-1}{M^{2}e^{R}2^{-M}}\right) .
\]
\encor

Note that although $\tilde\varrho^{-1}$ does depend on $R$ and $M$, as $M$ becomes 
large it converges to $e^{R}$. 
Observe that if we truncate a prior on $\reals^{d}$ to be on $\ball_{R}(\bzero)$,
by making $R$ and $M$ sufficiently large, the Wasserstein distance between $\post$
and the \pass posterior approximation $\apost$ can be made arbitarily small.
Similar results could be shown for other GLM likelihoods.

\section{Experiments}
\label{sec:experiments}

In our experiments, we focus on logistic regression, a particularly popular GLM example. Code is available at \url{https://bitbucket.org/jhhuggins/pass-glm}.
As discussed in \cref{sec:pass}, we choose~${M=2}$ and call our algorithm \passlr. 
Empirically, we observe that~${M=2}$ offers
a high-quality approximation of $\glmMap$ on the interval $[-4,4]$ (\cref{fig:2nd-degree-approx}).
In fact~${\sup_{s \in [-4,4]}|\glmMapM[2](s) - \glmMap(s)| < 0.069}$. 
Moreover, we observe that for many datasets, the inner products $\yn\xn \cdot \map$
tend to be concentrated within $[-4,4]$, and therefore 
 a high-quality approximation on this range is sufficient
for our analysis. In particular, \cref{fig:mode-inner-prods} shows histograms of $\yn\xn \cdot \map$ for four datasets
from our experiments.
In all but one case, over 98\% of the data points satisfy~${|\yn\xn \cdot \map| \le 4}$.
In the remaining dataset (\codrna), only $\sim$80\% of the data satisfy this condition,
and this is the dataset for which \passlr performed most poorly (cf.~\cref{cor:approx-post-logistic-simplified}).
We use a $\distNorm(\bzero, \sigma_{0}^{2}I)$ prior with $\sigma_{0}^{2} = 4$. 
Since we use a second order likelihood approximation, the \passlr posterior is Gaussian.
Hence we can calculate its mean and covariance in closed form.

\begin{figure}[tb]
\begin{center}
\begin{subfigure}[b]{0.247\textwidth} 
    \includegraphics[trim={0 40 0 0},clip,width=\textwidth]{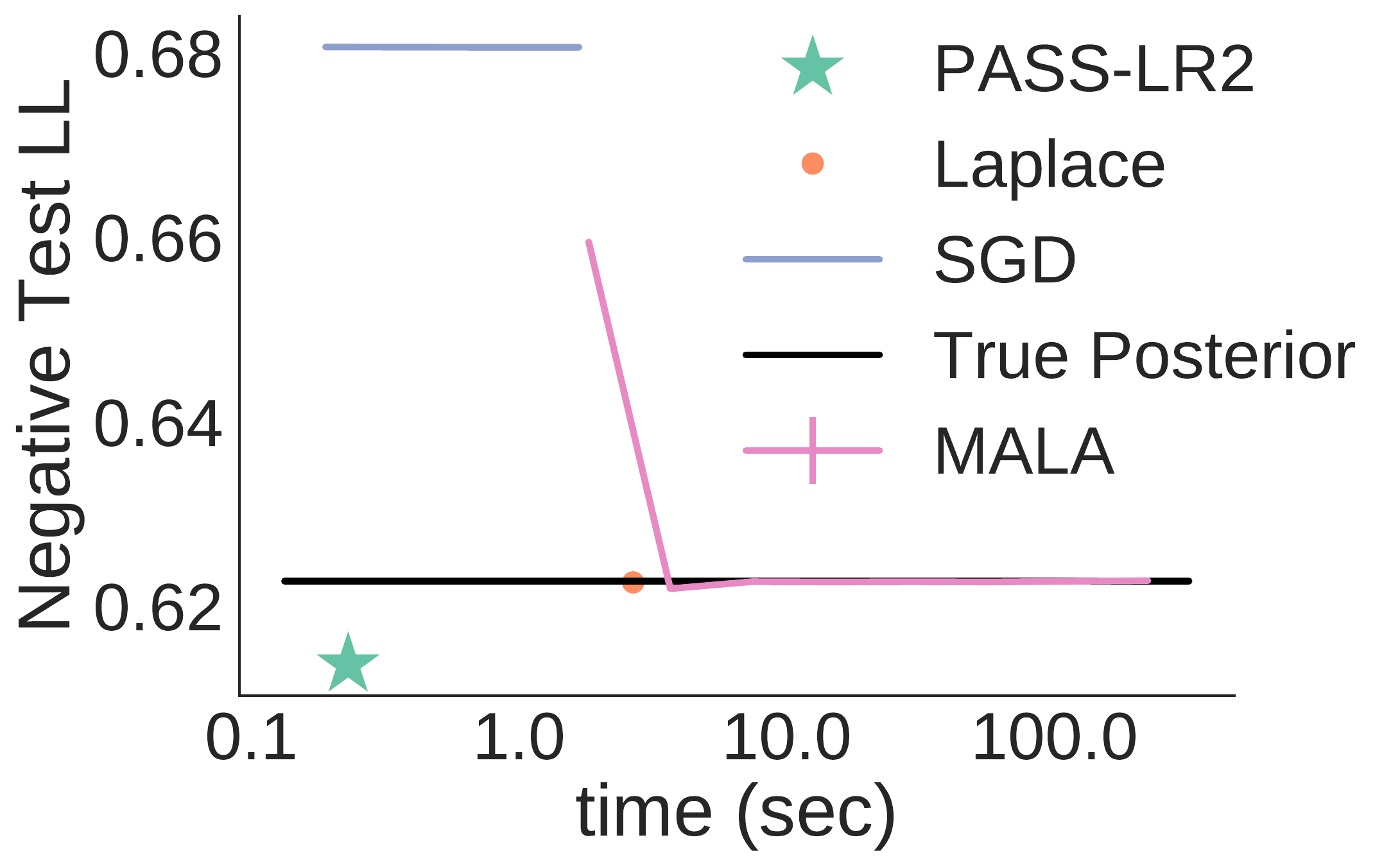} \\
    \includegraphics[trim={0 40 0 0},clip,width=.9\textwidth]{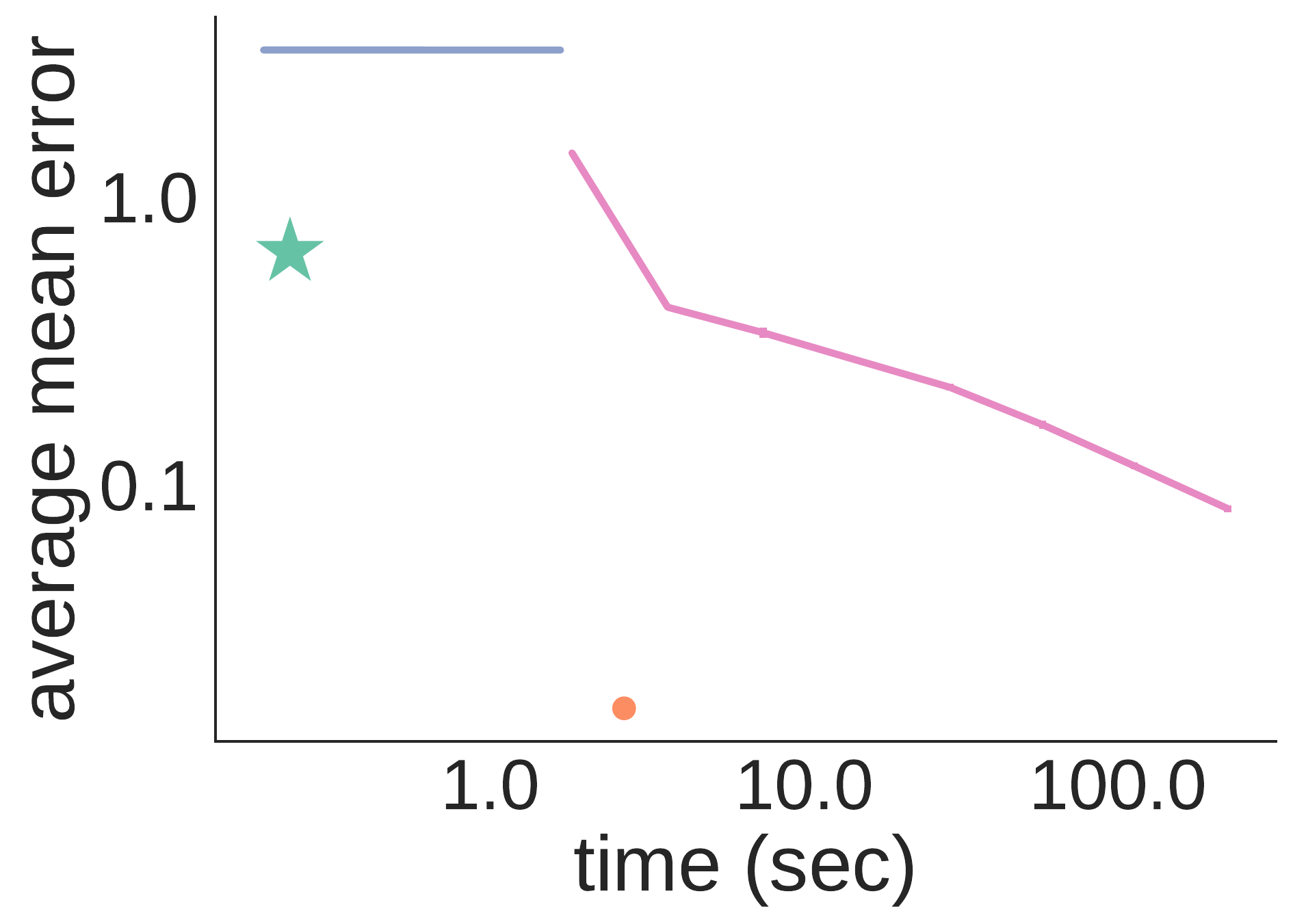} \\
    \includegraphics[width=.9\textwidth]{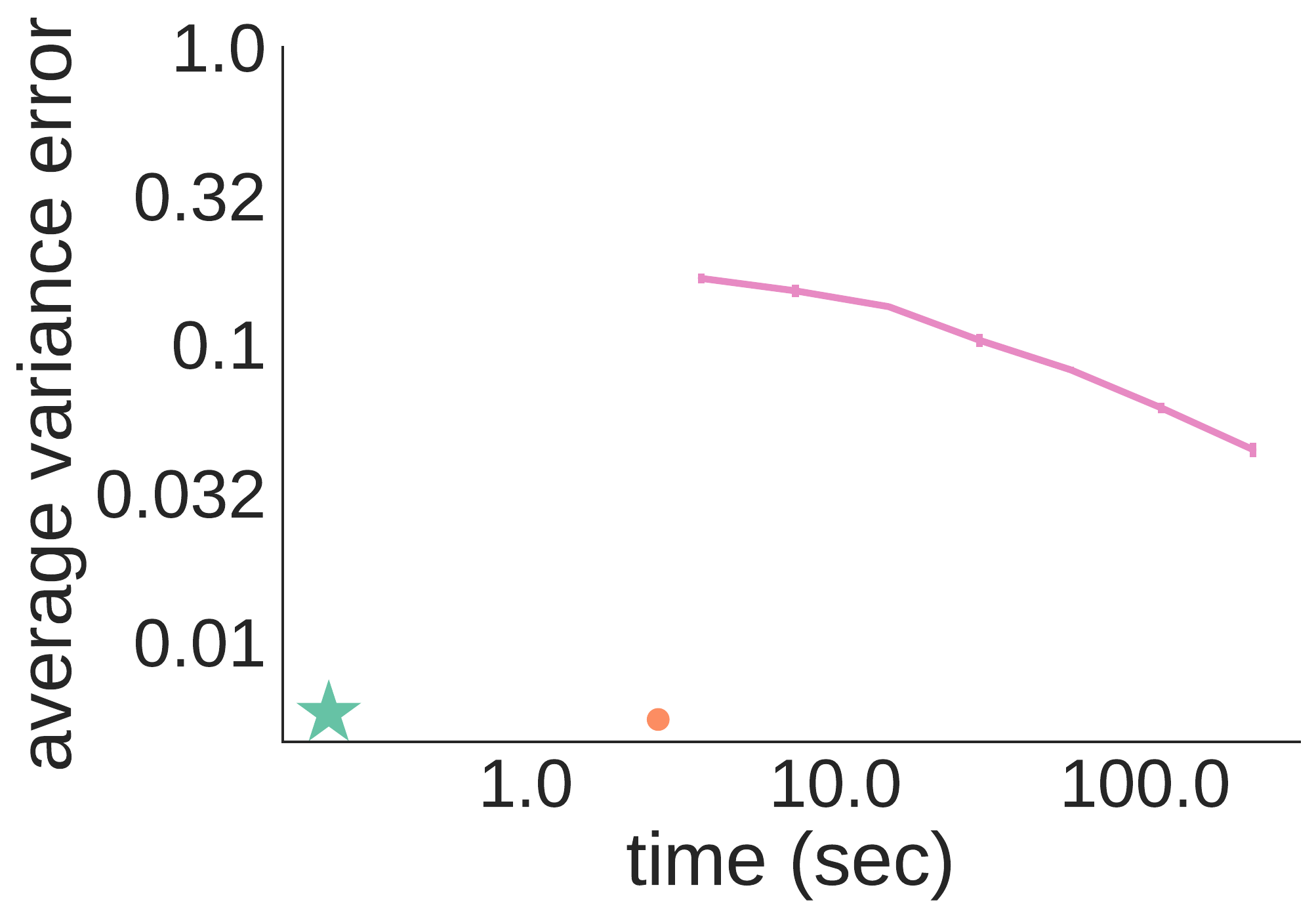} 
    \caption{\webspam}
    \label{fig:webspam-results}
\end{subfigure}  
\begin{subfigure}[b]{0.209\textwidth}
    \includegraphics[trim={40 40 0 0},clip,width=\textwidth]{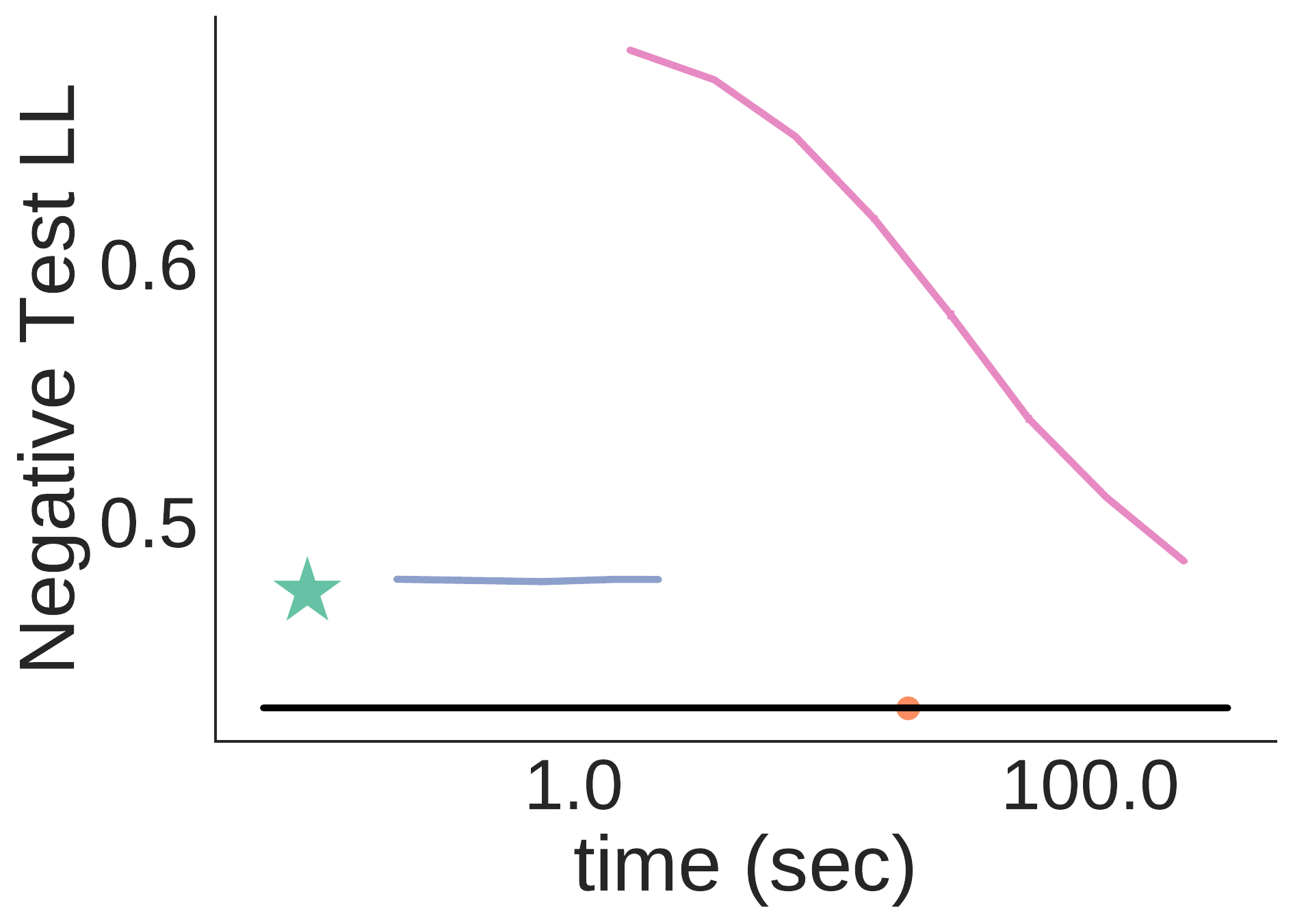} \\
    \includegraphics[trim={40 40 0 0},clip,width=\textwidth]{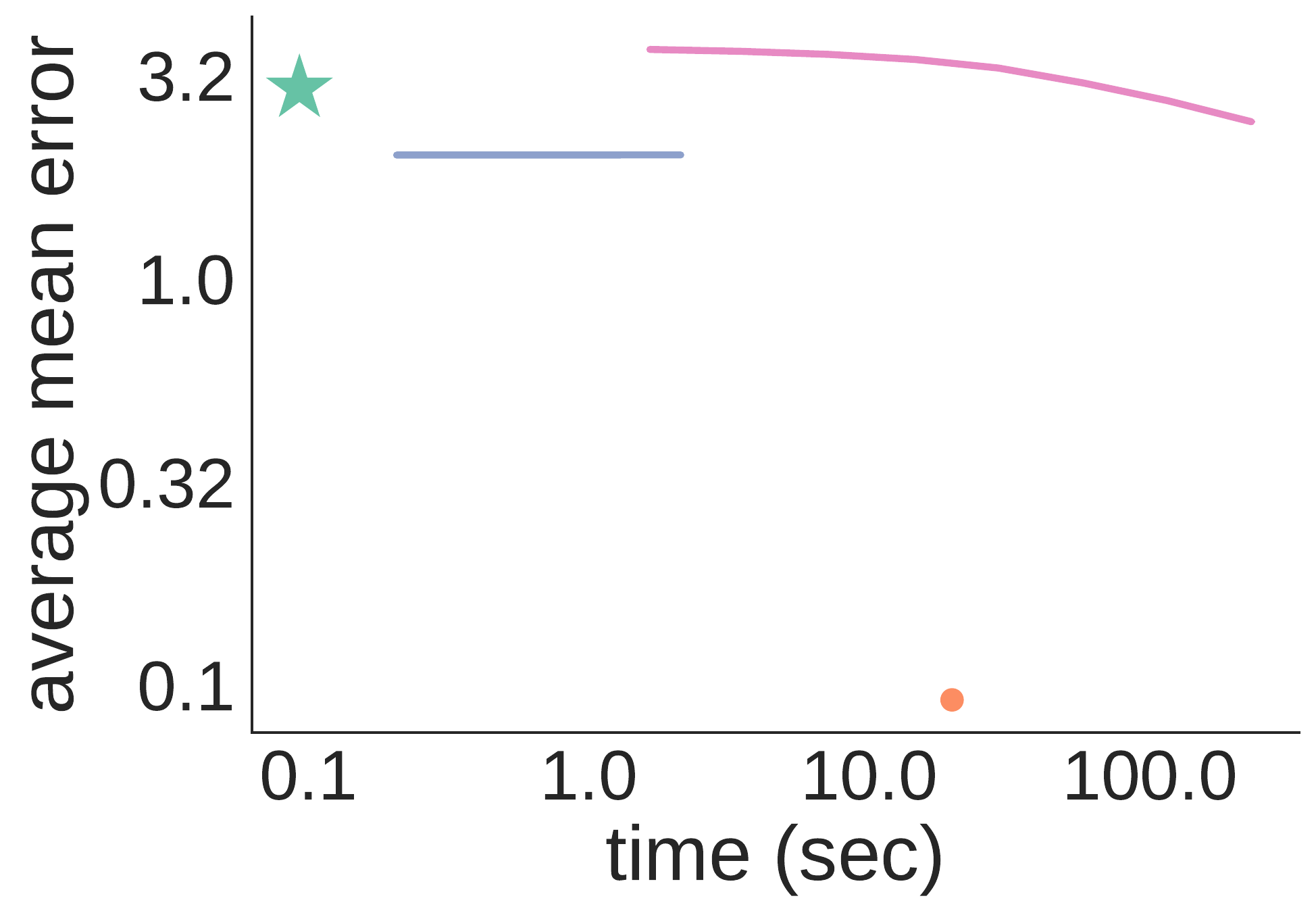} \\
    \includegraphics[trim={40 0 0 0},clip,width=\textwidth]{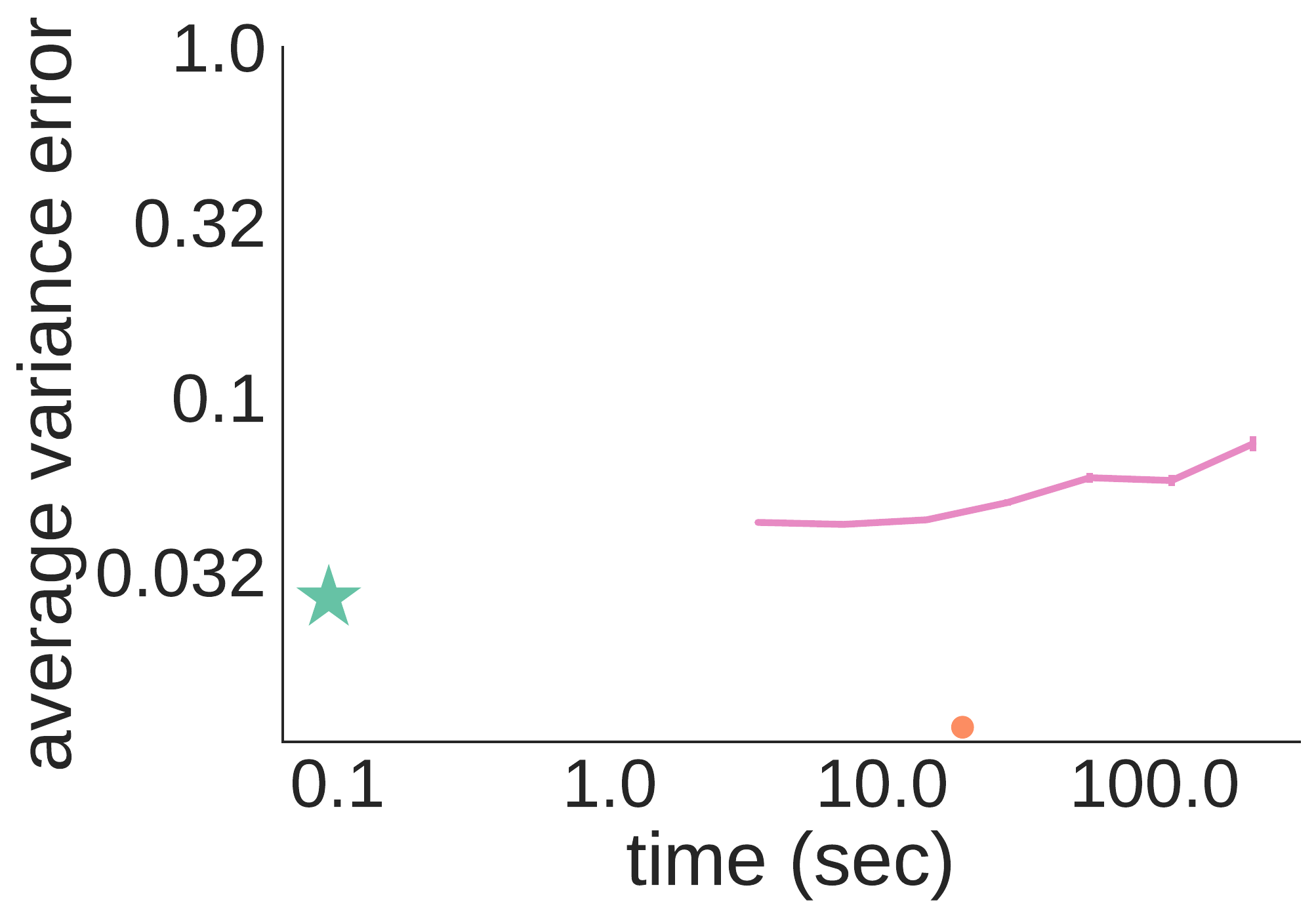} 
    \caption{\covtype}
    \label{fig:covtype-results}
\end{subfigure} 
\begin{subfigure}[b]{0.209\textwidth} 
    \includegraphics[trim={40 40 0 0},clip,width=\textwidth]{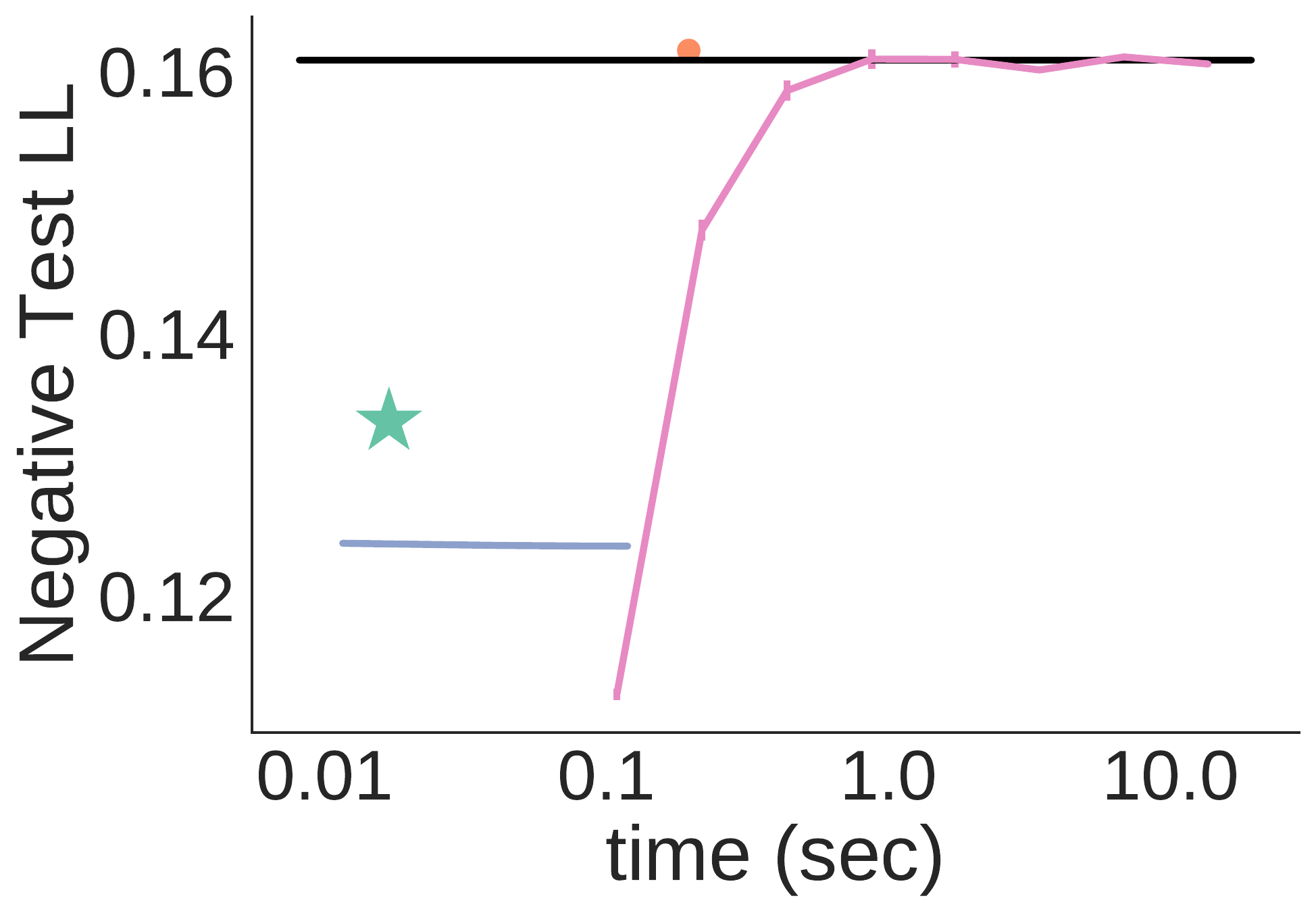} 
    \includegraphics[trim={40 40 0 0},clip,width=\textwidth]{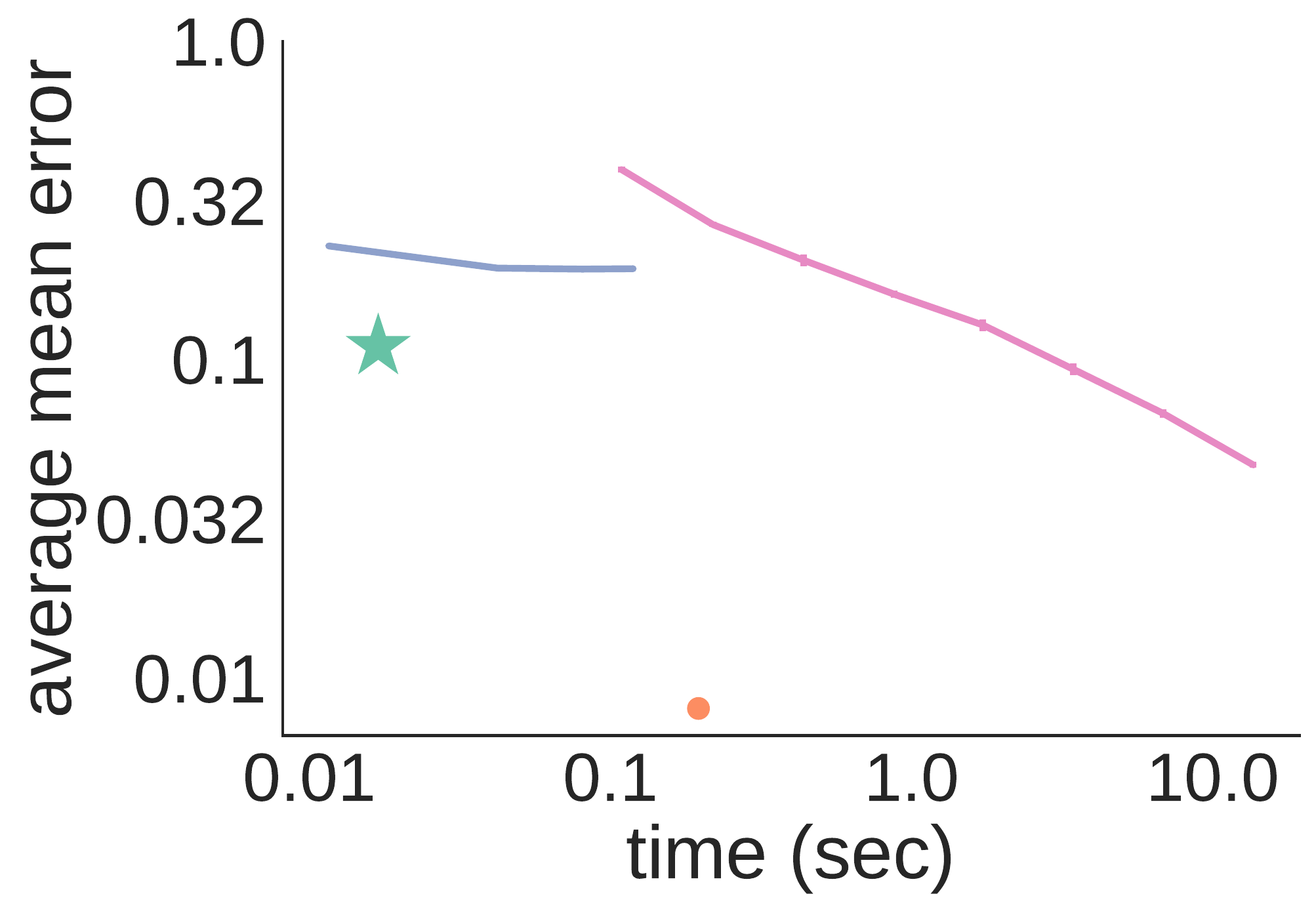} 
    \includegraphics[trim={40 0 0 0},clip,width=\textwidth]{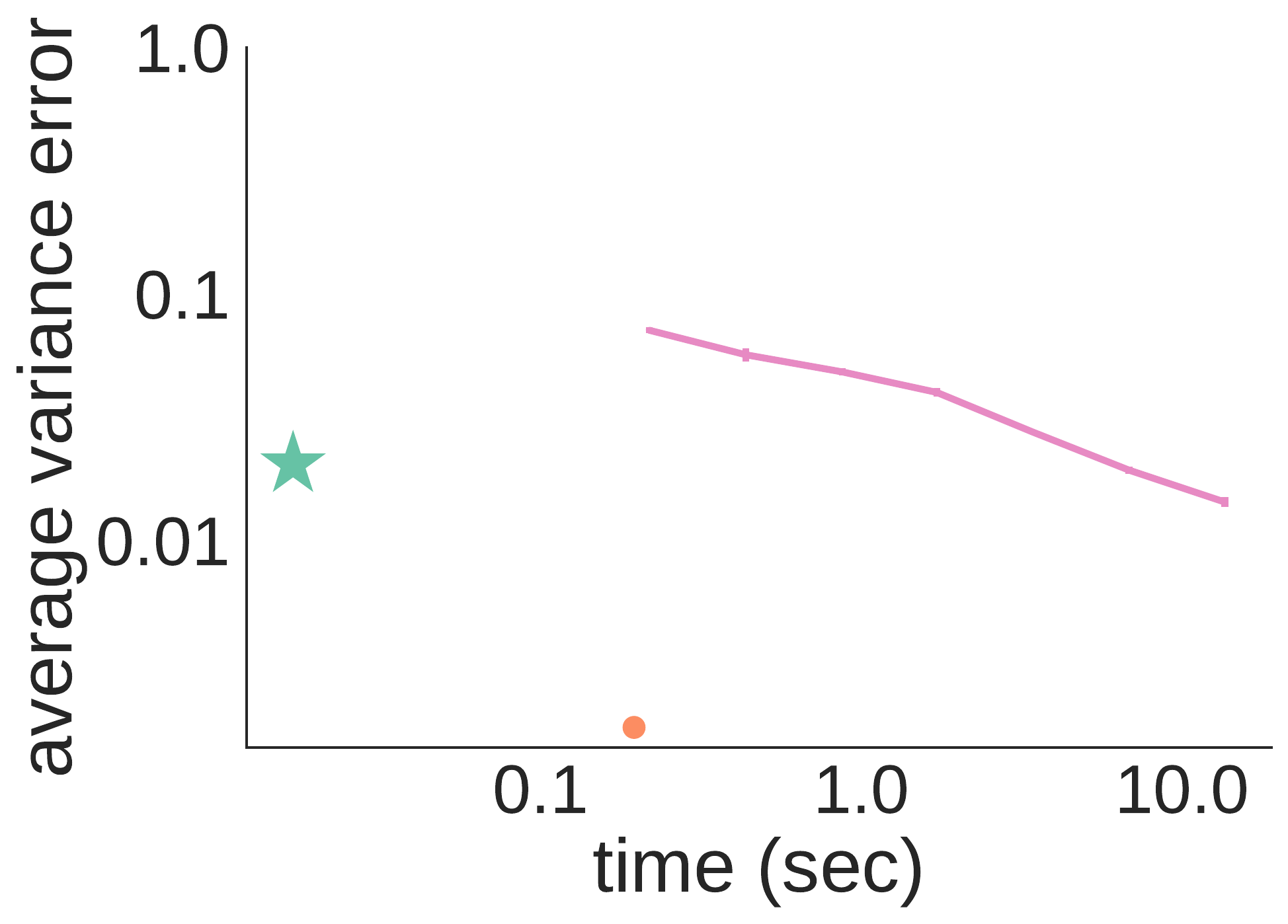} 
    \caption{\chemre}
    \label{fig:chemreact-results}
\end{subfigure}  
\vspace{1em}
\begin{subfigure}[b]{0.209\textwidth}
    \includegraphics[trim={40 40 0 0},clip,width=\textwidth]{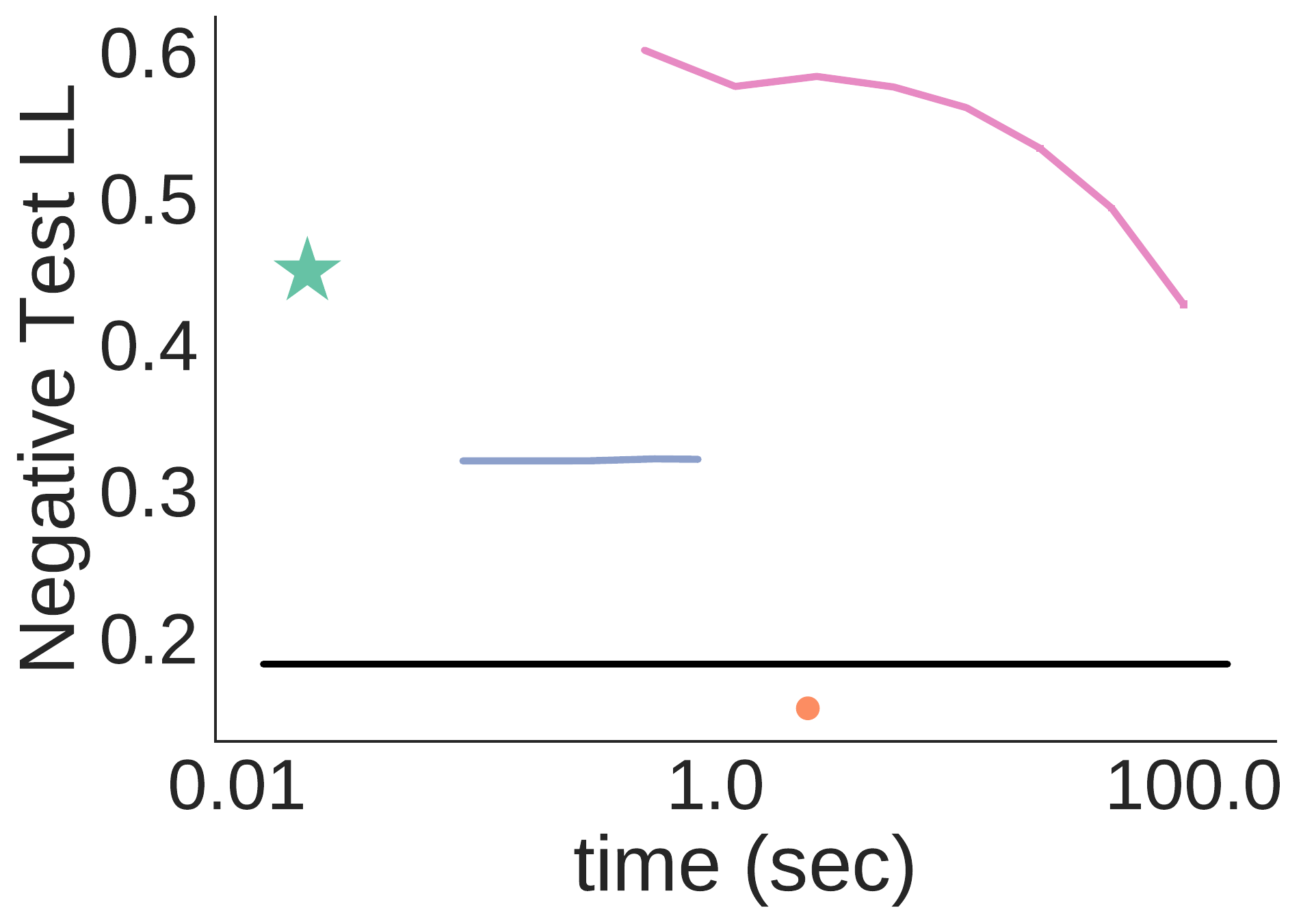} 
    \includegraphics[trim={40 40 0 0},clip,width=\textwidth]{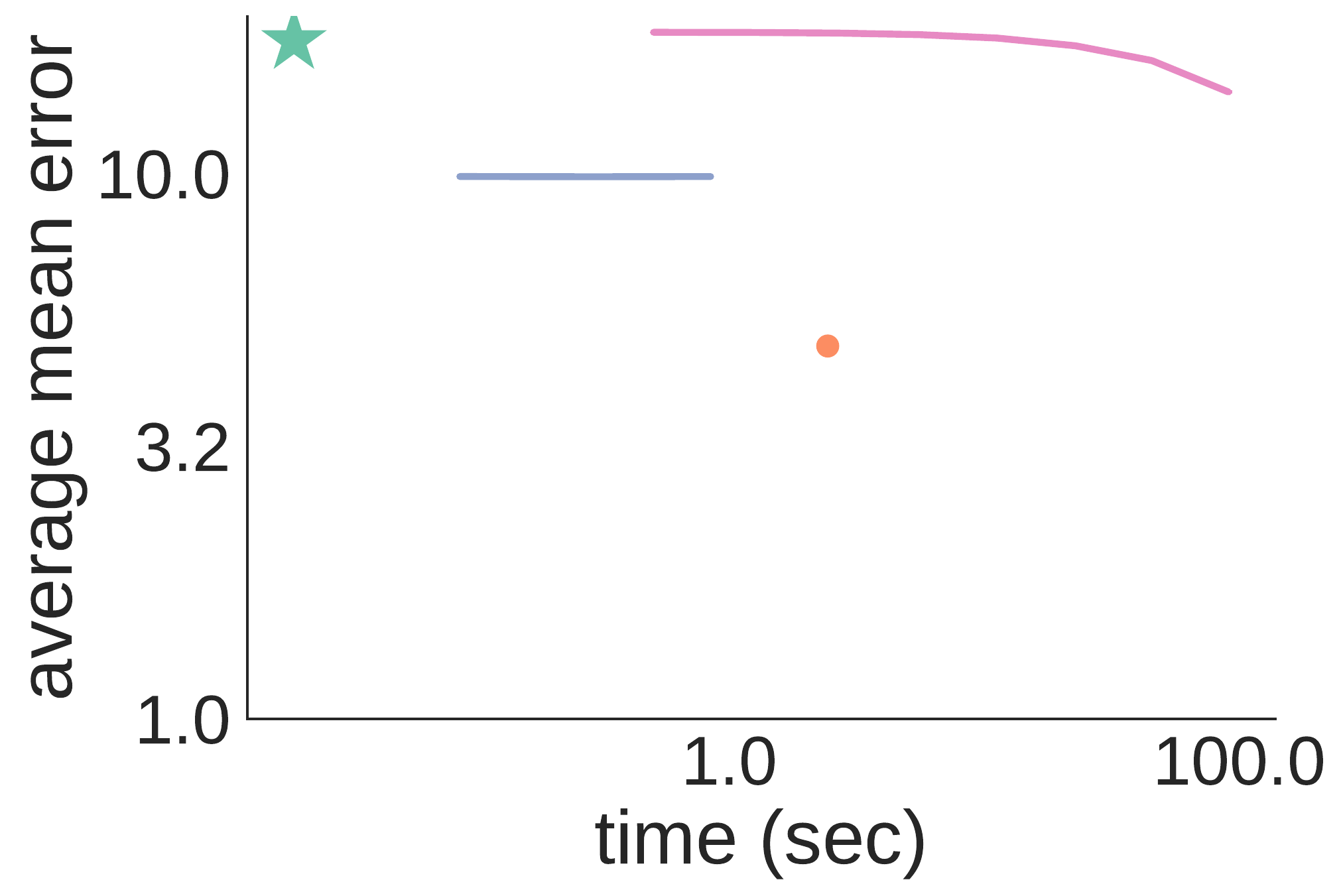} 
    \includegraphics[trim={40 0 0 0},clip,width=\textwidth]{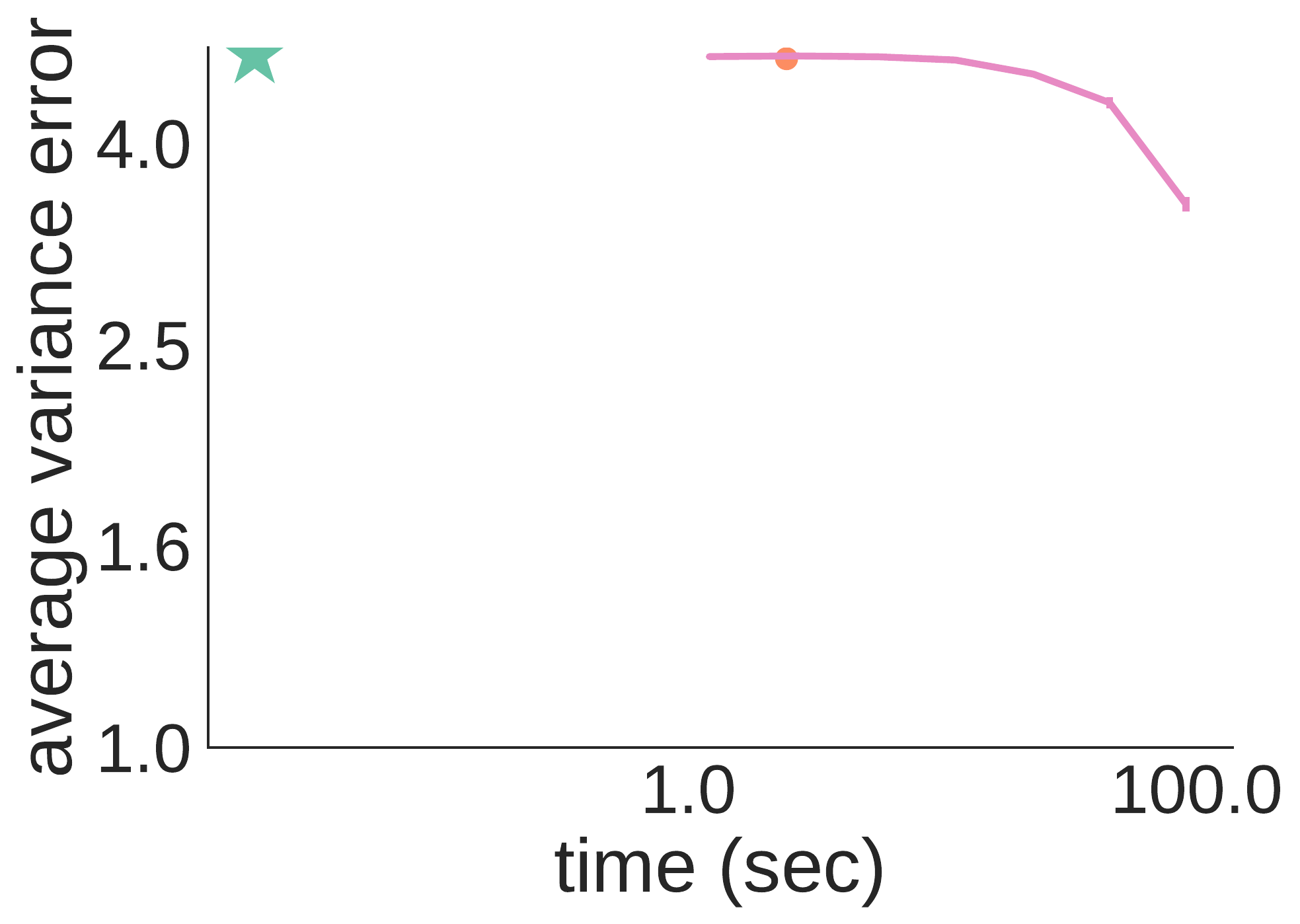} 
    \caption{\codrna}
    \label{fig:codrna-results}
\end{subfigure} 
\end{center}
\vspace{-.75em}
\caption{Batch inference results. In all metrics smaller is better.}
\vspace{-.5em}
\label{fig:results}
\end{figure}

\subsection{Large dataset experiments} \label{sec:compare_bayes}

In order to compare \passlr to other approximate Bayesian methods, we first restrict our attention to datasets with fewer than 1 million data points.
We compare to the Laplace approximation and the adaptive Metropolis-adjusted Langevin algorithm (MALA).
We also compare to stochastic gradient descent (SGD) although SGD provides only a point estimate and no approximate posterior.
In all experiments, no method performs as well as \passlr given the same (or less) running
time.

\textbf{Datasets.} The \chemre dataset consists of $N=$ 26,733 chemicals, each with ${d=100}$ properties. 
The goal is to predict whether each chemical is reactive. 
The \webspam corpus consists of $N=$ 350,000 web pages and the covariates consist of the ${d=127}$ 
features that each appear in at least 25 documents. 
The cover type (\covtype) dataset consists of $N=$ 581,012 cartographic observations with ${d=54}$ features.
The task is to predict the type of trees that are present at each observation location. 
The \codrna dataset consists of $N=$ 488,565 and ${d=8}$ RNA-related features.
The task is to predict whether the sequences are non-coding RNA.

\cref{fig:results} shows average errors of the posterior mean 
and variance estimates as well as  negative test log-likelihood for each method versus the time required to run the method. 
SGD was run for between 1 and 20 epochs.
The true posterior was estimated by running three chains of adaptive MALA for 50,000 iterations, which
produced Gelman-Rubin statistics well below 1.1 for all datasets.

\textbf{Speed.} For all four datasets, \passlr was an order of magnitude faster than SGD and 2--3 orders of magnitude
faster than the Laplace approximation. 
\textbf{Mean and variance estimates.} For \chemre, \webspam, and \covtype, \passlr was superior to or competitive with SGD, 
with MALA taking 10--100x longer to produce comparable results. 
Laplace again outperformed all other methods. 
Critically, on all datasets the \passlr variance estimates were competitive with Laplace and MALA. 
\textbf{Test log-likelihood.} For \chemre and \webspam, \passlr produced results competitive with all other methods. 
MALA took 10--100x longer to produce comparable results. 
For \covtype, \passlr was competitive with SGD but took a tenth of the time, and 
MALA took 1000x longer for comparable results. 
Laplace outperformed all other methods, but was orders of magnitude slower than \passlr. 
\codrna was the only dataset where \passlr performed poorly. 
However, this performance was expected based on the $\yn\xn \cdot \map$ histogram (\cref{fig:2nd-degree-approx}).

\subsection{Very large dataset experiments using streaming and distributed \pass}

\begin{figure}[tb]
\begin{center}
\begin{subfigure}[b]{.3\textwidth}
    \includegraphics[width=\textwidth]{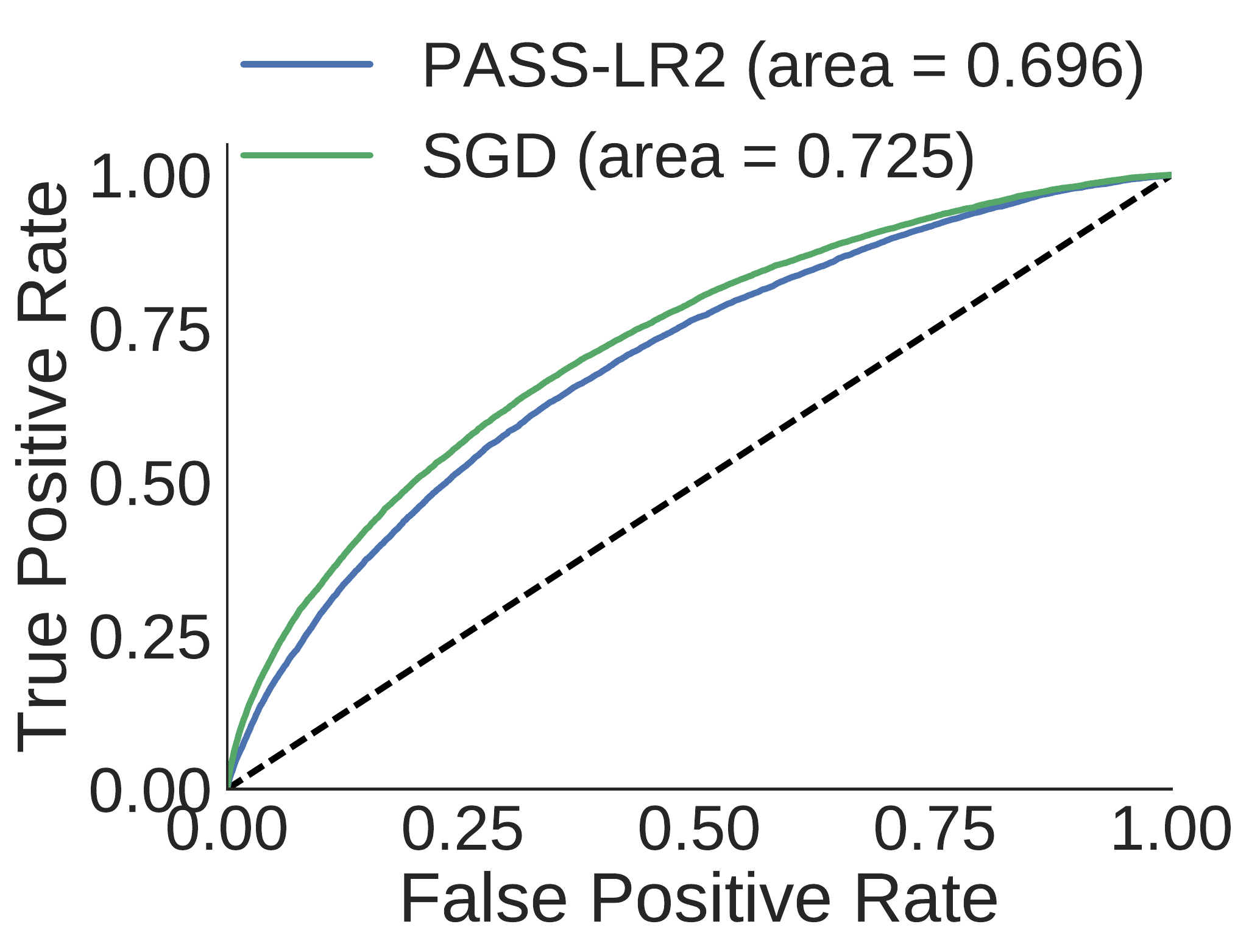}
    \caption{}
    \label{fig:criteo-roc-streaming}
\end{subfigure}
\hspace{2em}
\begin{subfigure}[b]{.3\textwidth}
    \includegraphics[width=\textwidth]{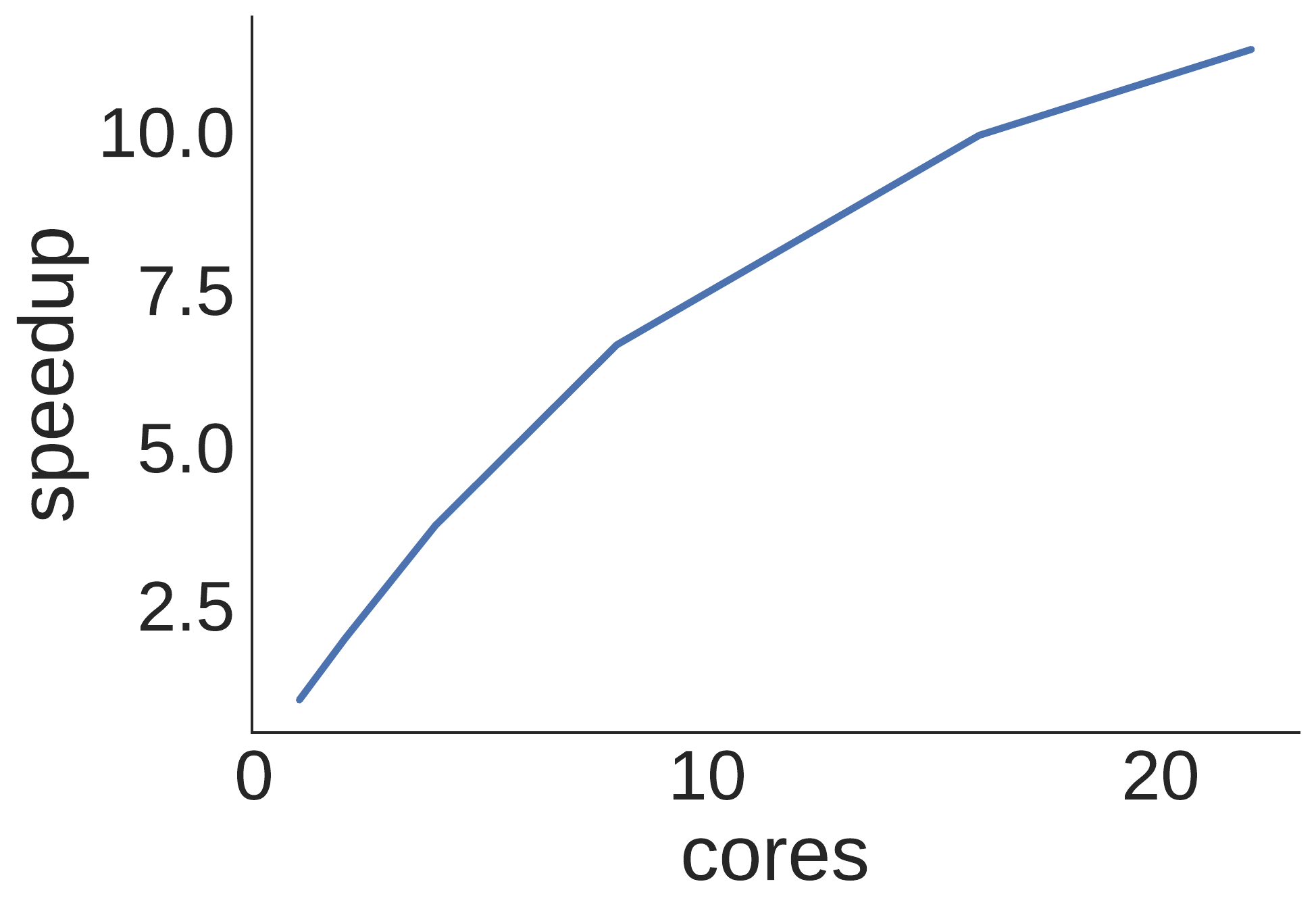}
    \caption{}
    \label{fig:criteo-distributed-scaling}
\end{subfigure}
\vspace{.5em}
\caption{\textbf{(a)} ROC curves for streaming inference on 40 million \criteo data points.
SGD and \passlr had negative test log-likelihoods of, respectively, 0.07 and 0.045. 
\textbf{(b)} Cores vs.\ speedup (compared to one core) for parallelization experiment on 6 million examples from the \criteo dataset. 
}
\vspace{-.75em}
\label{fig:streaming-and-distributed-results}
\end{center}
\end{figure}

We next test \passlr, which is streaming without requiring any modifications,
on a subset of 40 million data points from the Criteo terabyte ad click prediction dataset (\criteo).
The covariates are 13 integer-valued features and 26 categorical features.
After one-hot encoding, on the subset of the data we considered, $d \approx$ 3 million. 
For tractability we used sparse random projections~\citep{Li:2006} to reduce the dimensionality
to 20,000.
At this scale, comparing to the other fully Bayesian methods from \cref{sec:compare_bayes} was infeasible; we compare only to the predictions and point estimates from SGD.
\passlr performs slightly worse than SGD in AUC (\cref{fig:criteo-roc-streaming}), but outperforms SGD in negative test 
log-likelihood (0.07 for SGD, 0.045 for \passlr).
Since \passlr estimates a full covariance, it was about 10x slower than SGD. 
A promising approach to speeding up and reducing memory usage of \passlr would be
 to use a low-rank approximation to the second-order moments. 
 
To validate the efficiency of distributed computation with \passlr, we compared running times on
6M examples with dimensionality reduced to 1,000 when using 1--22 cores.
As shown in \cref{fig:criteo-distributed-scaling}, the speed-up is close to optimal: $K$ cores 
produces a speedup of about $K/2$ (baseline 3 minutes using 1 core). 
We used Ray to implement the distributed version of \passlr~\citep{nishihara2017}.\footnote{\url{https://github.com/ray-project/ray}} 

\section{Discussion}

We have presented \pass, a novel framework for scalable parameter estimation and Bayesian inference in generalized linear models. 
Our theoretical results provide guarantees on the quality of point estimates as well as approximate posteriors derived from \pass. 
We validated our approach empirically with logistic regression and a quadratic approximation. We showed competitive performance on a variety of real-world data, scaling to 40 million examples with 20,000 covariates, and trivial distributed computation with no compounding of approximation error.

There a number of important directions for future work. The first is to use randomization methods
along the lines of random projections and random feature mappings~\citep{Li:2006,Rahimi:2009tm} 
to scale to larger $M$ and $d$. 
We conjecture that the use of randomization will allow experimentation with other GLMs for which quadratic approximations are insufficient.

\subsubsection*{Acknowledgments}
{\small JHH and TB are supported in part by ONR grant N00014-17-1-2072, ONR
MURI grant N00014-11-1-0688, and a Google Faculty Research Award. RPA
is supported by NSF IIS-1421780 and the Alfred P. Sloan Foundation.}

\opt{arxiv}{%

\appendix

\numberwithin{equation}{section}
\numberwithin{figure}{section}

\section{General Derivation of PASS-GLM}
\label{app:pass-glm-general}

We can generalize the setup described in \cref{sec:pass} to cover a wide range of GLMs by assuming the log-likelihood is of the form 
\[
\log p(\yn[] \given \xn[], \param) 
= \sum_{k=1}^{K}\yn[]^{\alpha_{k}}\glmMapM[(k)](\yn[]^{\beta_{k}}\xn[] \cdot \param - a_{k}\yn[]), \label{eq:glm-general-form}
\]
where typically $\alpha_{k}, \beta_{k},a_{k} \in \{0,1\}$. 
We consider the $K=1$ case and drop the $k$ subscripts since the extension to $K > 1$ is trivial and serves only to introduce extra notational clutter. 
Letting $\glmMapM(s) = \sum_{m=0}^{M}\polyMapCoeffsM s^{m}$ be the order $M$ polynomial approximation to $\glmMap(s) = \glmMapM[(1)](s)$, we have that
\[
\log p(\yn[] \given \xn[], \param) 
&\approx \yn[]^{\alpha}\glmMapM(\yn[]^{\beta}\xn[] \cdot \param - a\yn[])  \\
&=  \yn[]^{\alpha}\sum_{m=0}^{M}  \polyMapCoeffsM(\yn[]^{\beta}\xn[] \cdot \param - a\yn[])^{m} \\
&= \yn[]^{\alpha} \sum_{m=0}^{M}  \polyMapCoeffsM\sum_{i=0}^{m}{m \choose i}(\yn[]^{\beta}\xn[] \cdot \param)^{i}(-a\yn[])^{m-i} \\
&=  \sum_{i=0}^{M}(\yn[]^{\beta}\xn[] \cdot \param)^{i}\yn[]^{\alpha}\sum_{m=i}^{M}  \polyMapCoeffsM{m \choose i}(-a\yn[])^{m-i} \\
&= \sum_{i=0}^{M}\sum_{\substack{\bk \in \nats^{d} \\ \sum_{j} k_{j} = i}} a'(\bk, i, M, \yn[])  \xn[]^{\bk}\param^{\bk},
\]
where $a'(\bk, \bar k, M, \yn[]) \defined \yn[]^{\alpha+i \beta}{ \bar k \choose \bk } \sum_{m=i}^{M}\polyMapCoeffsM {m \choose \bar k}(-a\yn[])^{m-\bar k}$. 
Thus, we have an exponential family model with 
\[
\suffstat(\xn[], \yn[]) &= \left(a'\left(\bk, {\textstyle\sum_{j}}k_{j}, M, \yn[]\right) \xn[]^{\bk}\right)_{\bk} &
\text{and}&&
\reparam(\param) &= (\param^{\bk})_{\bk},
\]
where $\bk$ is taken over all $\bk \in \nats^{d}$ such that $\sum_{j}k_{j} \le M$. 

The following examples show how a variety of GLM models fit into our framework. 
Throughout, let $s = \xn \cdot \param$. 

\begin{exa}[Robust regression] \label{ex:robust-regression}
For robust regression, $\obsspace = \reals$ and the log-likelihood is in the form $\glmMap(s - \yn[])$, where $\glmMap$ is a choice of ``distance'' function. 
For example, we could use either the Laplace likelihood 
\[
\laplaceMap(s - \yn[]) \defined -\frac{|s - \yn[]|}{b},
\]
the Cauchy likelihood 
\[
\cauchyMap(s - \yn[]) \defined -\ln\left(1 + \frac{(s - \yn[])^{2}}{b^{2}}\right),
\]
the negative Huber loss
\[
\huberMap(s - \yn[]) \defined 
\begin{cases}
-\frac{1}{2}(s - \yn[])^{2} & |s - \yn[]| \le b \\
-b|s - \yn[]| + \frac{1}{2}b^{2} & \text{otherwise},
\end{cases}
\]
or the negative smoothed Huber loss
\[
\smHuberMap(s - \yn[]) \defined -b^{2}\left(\sqrt{1 + \frac{(s - \yn[])^{2}}{b^{2}}} - 1\right),
\]
where in each case $b$ serves as a scale parameter. 
\end{exa}

\begin{exa}[Poisson regression]
For Poisson regression, $\obsspace = \nats$, and the log-likelihood is 
$\yn[]s - e^{s}$, so $\glmMapM[(1)](s) = s$, $\glmMapM[(2)](s) = -e^{s}$, $\alpha_{1} = 1$,
and $\beta_{1} = a_{1} = \alpha_{2} = \beta_{2} = a_{2} = 0$. 
\end{exa}

\begin{exa}[Gamma regression] \label{ex:gamma-regression}
For gamma regression, $\obsspace = \reals_{+}$, and the log-likelihood is $- \nu s - \nu \yn[] e^{-s} + c(\yn[], \nu)$ if using the log link,
where $\nu$ is a scale parameter. 
We can ignore the $c(\yn[], \nu)$ term since it does not depend on $\param$. 
Thus, $\glmMapM[(1)](s) = -\nu s$, $\glmMapM[(2)](s) = -\nu e^{-s}$, $\alpha_{2} = 1$,
and $\beta_{1} = a_{1} = \alpha_{1} = \beta_{2} = a_{2} = 0$.
\end{exa}

\begin{exa}[Probit regression]
For probit regression, $\obsspace = \{0, 1\}$, and the log-likelihood is 
\[
\begin{cases}
\ln(1 - \Phi(s)) & z = 0 \\
\ln(\Phi(s)) & z = 1
\end{cases},
\]
where $\Phi$ denotes the standard normal CDF. 
Thus, $\glmMapM[(1)](s) = \ln(1 - \Phi(s))$, $\glmMapM[(2)](s) = \ln(\Phi(s)) - \ln(1 - \Phi(s))$, $\alpha_{2} = 1$,
and $\beta_{1} = a_{1} = \alpha_{1} = \beta_{2} = a_{2} = 0$.
\end{exa}

\section{Chebyshev Approximation Results}
\label{app:chebyshev}

We begin by summarizing some standard results on the approximation accuracy of Chebyshev polynomials. 
Let $\glmMap : [-1,1] \to \reals$ be a continuous function, and let $\glmMapM$ be the $M$-th order Chebyshev approximation to $\glmMap$. 
Let $\|f\|_{\infty} \defined \sup_{s} |f(s)|$ be the $L^{\infty}$ norm of a function $f$; let $\comps$ denote the set of complex numbers;
and let $|z|$ be the absolute value of $z \in \comps$. 

\bnthm[{\citet[Theorem 5.14]{Mason:2003}}]
If $\glmMap$ has $k+1$ continuous derivatives, then $\|\glmMap - \glmMapM\|_{\infty} = O(M^{-k})$. 
\enthm

\bnthm[{\citet[Theorem 5.16]{Mason:2003}}] \label{thm:chebyshev-approx}
If $\glmMap$ can be extended to an analytic function on $E_{r} \defined \theset{z \in \comps  : |z + \sqrt{z^{2} -1}| = r}$ 
for $r > 1$ and $C \defined \sup_{z \in E_{r}} |\glmMap(z)|$, then
\[
\|\glmMap - \glmMapM\|_{\infty} \le \frac{C}{r-1}r^{-M}.
\]
\enthm

Chebyshev polynomials also provide a uniformly good approximation of the derivative of the function they
are used to approximate. 

\bnthm \label{thm:chebyshev-deriv-approx}
If $\glmMap$ can be extended to an analytic function on $E_{r}$
for $r > 1$ and $C \defined \sup_{z \in E_{r}} |\glmMap(z)|$, then
\[
\|\glmMap' - \glmMapM'\|_{\infty} \le C r^{-M} \frac{r+1}{(r-1)^{4}}\left[M^{2}r(r+1) + M(2r^{2} + r + 1) + r(r+1)\right] =: B(C, r, M)
\]
\enthm
\bprf
The proof follows the same structure as that for Theorem 5.16 in \citet{Mason:2003}.
For Chebyshev polynomials, $\basemeasure(\dee s) = \frac{2}{\pi}(1-s^{2})^{-1/2}\dee s$. 
Note that $\glmMap(s) = \sum_{m=0}^{\infty}  (\int \glmMap\basis\dee\basemeasure)\basis(s)$
and hence $\glmMap'(s) = \sum_{m=0}^{\infty}  (\int \glmMap\basis\dee\basemeasure) \basis'(s)$. 
Since $\basis' = m U_{m-1}$, where $\{U_{m}\}_{m\ge 0}$ are the Chebyshev polynomials of the second kind,
\[
\glmMap'(s) - \glmMapM'(s)
&= \sum_{m=M+1}^{\infty}\frac{2m}{\pi}\int_{-1}^{1}(1-v^{2})^{-1/2}\glmMap(v)\basis(v) U_{m-1}(s) \dee v.
\]
Define the conformal mappings $s = \frac{1}{2}(\xi + \xi^{-1})$ and $v = \frac{1}{2}(\zeta + \zeta^{-1})$, and $\glmMap(v) =: \tilde\glmMap(\zeta) = \tilde\glmMap(\zeta^{-1})$.
By assumption, $|\tilde\glmMap(\zeta)| \le C$. 
Let $\mcC_{1}$ denote the complex unit circle and for $r \in \reals_{+}$, let $\mcC_{r} \defined r\mcC_{1}$. 
Using the conformal mappings, we have
\[
\lefteqn{\glmMap'(s) - \glmMapM'(s)} \\
&= \sum_{m=M+1}^{\infty}\frac{m}{4\ii\pi}\oint_{\mcC_{1}}\tilde\glmMap(\zeta)(\zeta^{m} + \zeta^{-m}) \frac{\xi^{m} - \xi^{-m}}{\xi - \xi^{-1}}\frac{\dee \zeta}{\zeta} \\
&= \sum_{m=M+1}^{\infty}\frac{m}{2\ii\pi}\oint_{\mcC_{r}}\tilde\glmMap(\zeta)\zeta^{-m}\frac{\xi^{m} - \xi^{-m}}{\xi - \xi^{-1}}\frac{\dee \zeta}{\zeta} \\
&= \frac{1}{2\ii\pi}\oint_{\mcC_{r}}\frac{\tilde\glmMap(\zeta)}{\xi - \xi^{-1}}\left(\frac{\xi^{M+1} \zeta^{-M-1}(1 + M + \xi\zeta^{-1})}{(\xi\zeta^{-1} - 1)^2}-\frac{\xi^{-M-1}\zeta^{-M-1} (1 + M + \xi^{-1}\zeta^{-1})}{(\zeta^{-1}  \xi^{-1} -1)^2}\right)\frac{\dee \zeta}{\zeta} \\
&\le \frac{C}{2\ii\pi}\oint_{\mcC_{r}}\frac{\xi\zeta^{-M-1}\xi^{-M-1}}{\xi^{2} - 1}\left(\frac{\xi^{2M+2} (1 + M + \xi\zeta^{-1})}{(\xi\zeta^{-1} - 1)^2}-\frac{(1 + M + \xi^{-1}\zeta^{-1})}{(\zeta^{-1}  \xi^{-1} -1)^2}\right)\frac{\dee \zeta}{\zeta}.
\]
Letting $\eta \defined \xi^{2}$ and $\psi \defined \xi^{-1}\zeta^{-1}$, the absolute value of the integrand is
\[
\lefteqn{\frac{|\psi|^{M+1}}{|\eta - 1|}\left|\frac{\eta^{M+1}(1 + M - \eta\psi)}{(\eta \psi - 1)^{2}} - \frac{1 + M - \psi}{(\psi - 1)^{2}}\right|} \\
&= r^{-M-1}\frac{|\eta \psi - 1|^{-2}|\psi - 1|^{-2}}{|\eta - 1|}\left|\eta^{M+1}(1 + M - \eta\psi)(\psi - 1)^{2}- (1 + M - \psi)(\eta \psi - 1)^{2}\right| \\
\begin{split}
&\le r^{-M-1}\frac{(r^{-1}-1)^{-4}}{|\eta - 1|}\Big[|\psi||\eta^{M+2}-1| + (M+1)|\eta^{M+1} - 1| +  2|\psi|^{2}|\eta^{M+1} - 1|  \\
&\hphantom{\le r^{-M-1}\frac{(r^{-1}-1)^{-4}}{|\eta - 1|}\Big[}~+ 2(M+1)|\psi||\eta^{M}-1| + |\psi|^{3}|\eta^{M}-1| + (M+1)|\phi|^{2}|\eta^{M-1}-1| \Big]
\end{split} \\
&\le \frac{r^{-M+3}}{(r-1)^{4}}\left[\frac{M+2}{r} + (M+1)^{2} + \frac{2(M+1)}{r^{2}} + \frac{2M(M+1)}{r} + \frac{M}{r^{3}} + \frac{M^{2} - 1}{r^{2}}\right] \\
&= r^{-M} \frac{r+1}{(r-1)^{4}}\left[M^{2}r(r+1) + M(2r^{2} + r + 1) + r(r+1)\right].
\]
The final inequality follows from the fact that for $k \in \nats$, 
\[
|\eta^{k} - 1|/|\eta - 1| = |\sin(k\arg(\eta))/\sin(\arg(\eta)| \le k.
\]
The result now follows. 
\eprf

Since $\logitMap$ is smooth, we can apply \cref{thm:chebyshev-approx,thm:chebyshev-deriv-approx} to obtain exponential convergence rates 
of the (derivative of the) Chebyshev approximation. 
The same is true in the Poisson and smoothed Huber regression cases. 

\bncor \label{cor:logit-chebyshev-error}
Fix $R > 0$. 
If $\glmMap(s) = \log(1 + e^{-R s})$, $s \in [-1,1]$, then for 
any $r \in (1, \pi/R  + \sqrt{\pi^{2}/R^{2} + 1})$, 
\[
\|\glmMap - \glmMapM\|_{\infty} \le \frac{C(r, R) }{(r-1)r^{M}} 
\qquad \text{and}  \qquad
\|\glmMap' - \glmMapM'\|_{\infty} \le  B(C(r, R), r, M),
\]
where $C(r, R) \defined \left|\log\left(1+ e^{-\frac{1}{2}R (r-r^{-1})\ii} \right)\right|$. 
\encor
\bprf
The function $e^{-R s}$ is entire while $\log$ is analytic except at $0$.
Thus, we must determine the minimum value of $r$ such that there exists $z \in E_{r}$ such that $1 + e^{-R z} = 0$. 
Taking $z = a + b \ii$, it must hold that $b \in \{ k\pi/R : k \in \ints \}$ since otherwise
$e^{-Rz}$ would contain an imaginary component. 
If $b = 2k\pi/R$ then $e^{-Rz} = e^{-Ra} > 0$, so this cannot be a solution to $1 + e^{-Rz} = 0$. 
However, taking $b = (2k + 1)\pi/R$ yields $1 - e^{-Ra} = 0 \implies a = 0$. 
Hence, $z = (2k+1)\pi\ii/R$ and thus
\[
|z + \sqrt{z^{2} - 1}| 
&= |\pi \ii/R + \sqrt{-(2k+1)^{2}\pi^{2}/R^{2} - 1}| \\
&= |(\pi/R + \sqrt{(2k+1)^{2}\pi^{2}/R^{2} + 1})\ii| \\
&= \pi/R + \sqrt{(2k+1)^{2}\pi^{2}/R^{2} + 1} \\
&\ge \pi/R + \sqrt{\pi^{2}/R^{2} + 1}.
\]
Thus we must choose $r < \pi/R + \sqrt{\pi^{2}/R^{2} + 1}$. 
For any such $r$, $|\glmMap(z)|$ is maximized along $E_{r}$ when $z = b \ii$, which implies 
$b = \frac{1}{2}(r-r^{-1})$ 
and hence $C = C(r, R)$. 
The two inequalities now follow from, respectively, \cref{thm:chebyshev-approx,thm:chebyshev-deriv-approx}.
\eprf
\bncor \label{cor:exp-chebyshev-error}
Fix $R > 0$. 
If $\glmMap(s) = e^{Rs}$, $s \in [-1,1]$, then for 
any $r > 1$, 
\[
\|\glmMap - \glmMapM\|_{\infty} &\le \frac{e^{\frac{1}{2}R (r+r^{-1})}}{(r-1)r^{M}} \\
\|\glmMap' - \glmMapM'\|_{\infty} &\le  B(e^{\frac{1}{2}R (r+r^{-1})}, r, M).
\]
\encor
\bprf
The proof is similar to that for \cref{cor:logit-chebyshev-error}.
The differences are as follows.
The function $e^{-R s}$ is entire, so we may choose any $r > 1$.  
For any such $r$, $|\glmMap(z)|$ is maximized along $E_{r}$ when $z$ is real, which implies 
$z = \frac{1}{2}(r+r^{-1})$ 
and hence $C = e^{\frac{1}{2}R (r+r^{-1})}$. 
\eprf

\bncor \label{cor:smoothed-huber-chebyshev-error}
Fix $R > 0$. 
If $\glmMap(s) = b^{2}\left(\sqrt{1 + \frac{R^{2}s^{2}}{b^{2}}} - 1\right)$, $s \in [-1,1]$, then for 
any $r \in (1, b/R + \sqrt{b^{2}/R^{2} + 1})$, 
\[
\|\glmMap - \glmMapM\|_{\infty} &\le \frac{b^{2}\sqrt{1 + \{(r^{2}+1)/(2rb)\}^{2}} - b^{2}}{r-1}r^{-M} \\
\|\glmMap' - \glmMapM'\|_{\infty} &\le  B\left( b^{2}\sqrt{1 + \{(r^{2}+1)/(2rb)\}^{2}} - b^{2}, r, M\right).
\]
\encor
\bprf
The proof is similar to that for \cref{cor:logit-chebyshev-error}.
The differences are as follows.
The square root function is analytic except at zero, so we must determine the minimum value of $r$
such that there exists $z \in E_{r}$ such that $1 + R^{2}z^{2}/b^{2} = 0$.
Solving, we find that $z = \ii b/R$. 
Thus, we have 
\[
|z + \sqrt{z^{2} - 1}| = b/R + \sqrt{b^{2}/R^{2} + 1}
\]
and so must choose $1 < r < b/R + \sqrt{b^{2}/R^{2} + 1}$. 
For any such $r$, $|\glmMap(z)|$ is maximized along $E_{r}$ when $z$ is real, which implies $z = \frac{r^{2} + 1}{2r}$
and hence $C = b^{2}\left(\sqrt{1 + \left(\frac{r^{2}+1}{2rb}\right)^{2}} - 1\right)$.
\eprf

\section{Approximation Theorems and Proofs}
\label{app:proofs}

\bnthm \label{thm:approx-map}
Let~${\ball_{r}(\param^{*}) \defined \theset{\param \in \paramspace \given \|\param - \param^{*}\|_{2} \le r}}$. 
Assume there exist parameters $\veps_{N}$ and $\varrho_{N}$ such that for all 
$\param \in \ball_{r_{N}}(\map)$, where $r_{N}^{2} \defined 4\veps_{N}/\varrho_{N}$, 
\benum[label=(\Alph*),wide,series=assumptions]
\item \label{asm:lliks-close} $|\llik(\param) - \allik(\param)| \le \veps_{N}$ and \qquad\qquad
\inlineitem \label{asm:log-post-strongly-concave} $-\log \post$ is $\varrho_{N}$-strongly convex.\footnote{A differentiable function~${f : \reals^{d} \to \reals}$ is~$\varrho$-strongly convex if for all~${\bv, \bw \in \reals^{d}}$, 
$f(\bv) \ge f(\bw) + \ip{\grad f(\bw)}{\bv - \bw} + (\varrho/2)\|\bv - \bw\|_{2}^{2}.$}
\eenum 
Furthermore, assume that for all $\param \in \paramspace$,
\benum[resume*=assumptions]
\item \label{asm:log-post-quasi-concave} $\log \post$ is strictly quasi-concave\footnote{An arbitrary function~${g : \reals^{d} \to \reals}$ is strictly quasi-concave if for all~${\bv, \bw \in \reals^{d}}$, ${\bv \ne \bw}$, and ${t \in (0,1)}$, 
${g(t\bv + (1-t)\bw) > \min\{ g(\bv), g(\bw) \}}$.}
and ~~~
\inlineitem \label{asm:allik-lb} $\allik(\param) \le \llik(\param) + \veps_{N}$.  %
\eenum
Then  $\|\map - \amap\|^{2}_{2} \le \frac{4\veps_{N}}{\varrho_{N}}$.%
\enthm

\brmk[Assumptions]
The error in the MAP estimate naturally depends on the error of the approximate log-likelihood (Assumption~\ref{asm:lliks-close})
as well as the flatness of the posterior (Assumption~\ref{asm:log-post-strongly-concave}).
In the latter case, if $\log \post$ is very flat, then even a small error from using $\allik$ in
place of $\llik$ could lead to a large error in the approximate MAP solution. 
However, the stronger assumptions, \ref{asm:lliks-close} and \ref{asm:log-post-strongly-concave},
need hold only near the MAP solution.
\ermk
\brmk[Strict quasi-concavity]
Requiring that $\log \post$ be only strictly quasi-concave (rather than strongly log-concave everywhere) substantially 
increases the applicability of the result.
For instance, it allows heavy-tailed priors (e.g., Cauchy)
as well as sparsity-inducing priors (e.g., Laplace/$L_{1}$ regularization).
\ermk

\bprf[Proof of \cref{thm:approx-map}]
An equivalent condition for $f$ to be strictly quasi-convex is that if $f(\bv) > f(\bw)$ then
$\ip{\grad f(\bw)}{\bv - \bw} > 0$~\citep[Theorem 21.14]{Simon:1994}.
We obtain the result by considering some $\param$ such that $\param \notin \ball_{r_{N}}(\map)$.
Since $\logpost \defined \log\post$ is strictly quasi-concave (by Assumption \ref{asm:log-post-quasi-concave}), 
if it has a global maximum it is unique (if it had two global maxima, this would immediately yield a contradiction).
By hypothesis $\map$ is such a global maximum.
Thus, $\logpost(\map) > \logpost(\param)$, which implies 
\[
\ip{\grad \logpost(\param)}{\map - \param} > 0. \label{eq:map-quasi-ineq}
\]
Now, fix $\param'$ such that $\param' \notin \ball_{r_{N}}(\map)$.
Let $r_{N}' \defined \|\param' - \map\|_{2} > r_{N}$ and $\param'' \defined \frac{r_{N}}{r_{N}'}\param' + \frac{r_{N}'-r_{N}}{r_{N}'}\map$, the projection of $\param'$ onto $\ball_{r_{N}}(\map)$.
Applying the fundamental theorem of calculus for line integrals on the linear path 
$\gamma[\param', \param'']$ from $\param'$ to $\param''$, parameterized as 
$\param(t) = t\param'' + (1-t)\param'$, we have
\[
\llik(\param'') - \llik(\param')
&= \int_{\gamma[\param', \param'']} \grad \logpost(\param) \cdot \dee \param \\
&= \int_{0}^{1} \grad \logpost(\param(t)) \cdot (\param'' - \param')\,\dee t \\
&=  \frac{r_{N}'-r_{N}}{r_{N}'}\int_{0}^{1} \grad \logpost(\param(t)) \cdot (\map - \param') \,\dee t \\
&=  \frac{r_{N}'-r_{N}}{r_{N}'}\int_{0}^{1} C(t) \grad \logpost(\param(t)) \cdot (\map - \param(t)) \dee t \\
&> 0,
\]
where $C(t) \defined \frac{r_{N}'}{r_{N}' - tr_{N}' + t r_{N}}$
and the inequality follows from \cref{eq:map-quasi-ineq}. 
Hence,
\[
\logpost(\param') < \logpost(\param'') \label{eq:param-proj-ineq}
\] 
and 
\[
\log\prior(\param') + \allik(\param')
&\le \log\prior(\param') +  \llik(\param') + \veps_{N} && \text{by Assumption \ref{asm:allik-lb}} \\
&< \log\prior(\param'') + \llik(\param'') + \veps_{N} && \text{by \cref{eq:param-proj-ineq}} \\
&\le \log\prior(\map) + \llik(\map) + \veps_{N} - \frac{\varrho_{N}r_{N}^{2}}{2} && \text{by Assumption \ref{asm:log-post-strongly-concave}} \\
&=  \log\prior(\map) + \llik(\map) - \veps_{N}  && \text{by definition of $r_{n}$} \\
&\le \log\prior(\map) + \allik(\map) && \text{by Assumption \ref{asm:lliks-close}.}
\]
So $\param'$ is not a global optimum of $\log\apost$ and hence $\amap \in \ball_{R_{N}}(\map)$. 
\eprf

We present a generalization of \cref{cor:logit-map-error}. 
Let $\|\bT\|_{op} \defined \sup_{\substack{\bv \in \reals^{d} \\ \|\bv\|_{2} = 1}} \|\bT[\bv]\|_{op}$ 
denote the operator norm of the tensor $\bT$ (with $\|\bT\|_{op} = \|\bT\|_{2}$ if $\bT$ is a matrix). 
Recall the Lipschitz operator bound property
\[
 \|\grad h(x)\|_{op} &= \sup_{y \ne x} \frac{\|h(x) - h(y)\|_{op}}{\|x - y\|_{2}}, \label{eq:op-lipschitz}
\]
which holds for any sufficiently smooth $h : \reals^{d} \to (\reals^{d})^{\otimes k}$. 
Recall also that for compatible operators $T$ and $T'$, $\|TT'\|_{op} \le \|T\|_{op}\|T'\|_{op}$. 

\bncor \label{cor:logit-map-error-general}
Assume the tensor defined by ${T_{ijk} \defined \sum_{n=1}^{N}x_{ni}x_{nj}x_{nk}}$ satisfies ${\|\bT\|_{op} \le LN/d^{2}}$.
For the logistic regression model, assume that ${\|\grad^{2}\llik(\map)^{-1}\|_{2} \le cd/N}$ 
and that ${\|\xn\|_{2} \le 1}$ for all $n=1,\dots,N$.
Let $\glmMapM$ be the order $M$ Chebyshev approximation to $\logitMap$ on $[-R,R]$ such that \cref{eq:bounding-assumption} holds. %
Let $\apost(\param)$ denote the posterior approximation obtained by using $\glmMapM$ 
with a strictly quasi-log concave prior. 
Let 
\[ 
\veps \defined \min_{r \in (1,\pi/R  + \sqrt{\pi^{2}/R^{2} + 1})} \left|\log\left(1+ e^{-\frac{1}{2}R (r-r^{-1})\ii} \right)\right|(r-1)^{-1}r^{-M}
\]
and ${\alpha^{*} \defined 1 + b - \sqrt{(b+1)^{2} - 1}}$, where $b \defined \frac{\veps L^{2}c^{3}}{54d}$.
If ${R - \|\map\|_{2} \ge 2\sqrt{\frac{cd \veps}{\alpha^{*}}}}$, then 
\[
\|\map - \amap\|_{2}^{2} 
\le \frac{4cd \veps}{\alpha^{*}} 
\le \frac{4}{27}c^{4}L^{2}\veps^{2} + 8cd\veps
\]
and \cref{cor:logit-map-error} follows from the upper bound ${\|\bT\|_{op} \le N}$ 
(using the assumption that $\|\xn\|_{2} \le 1$). 
\encor
\bprf
By \cref{cor:logit-chebyshev-error}, for all $s \in [-R, R]$, $|\logitMap(s) - \glmMapM(s)| \le \veps N$. 
It is easy to verify that $\max_{s \in \reals}|\logitMap'''(s)| = \frac{1}{6\sqrt{3}}$
and therefore $\|\grad^{3}\llik(\param)\|_{op} \le \frac{1}{6\sqrt{3}}\|\bT\|_{op} \le \frac{LN}{6\sqrt{3}d^{2}}$. 
Since by hypothesis ${\|(\grad^{2}\llik(\map))^{-1}\|_{2} \le cd/N}$,
$\llik(\map)$ is $N/(cd)$-strongly concave. 
We can write $\grad (\grad^{2}\llik)^{-1} = - (\grad^{2}\llik)^{-1} \grad^{3}\llik (\grad^{2}\llik)^{-1}$ if we treat
the first $(\grad^{2}\llik)^{-1}$ as a matrix to matrix operator, $\grad^{3}\llik$ as a vector to matrix operator, and the second 
$(\grad^{2}\llik)^{-1}$ as a vector to vector operator.  
Thus 
\[
\|\grad (\grad^{2}\llik)^{-1}(\param)\|_{op} 
\le  \|(\grad^{2}\llik)^{-1}(\param)\|_{op}^{2}\|\grad^{3}\llik(\param)\|_{op} 
\le \frac{c^{2}d^{2}}{N^{2}}\frac{LN}{6\sqrt{3} d^{2}}
= \frac{c^{2}L}{6\sqrt{3} N}.
\]
Using the triangle inequality and \cref{eq:op-lipschitz}, we have
\[
\|(\grad^{2}\llik)^{-1}(\param)\|_{op} 
&\le  \|(\grad^{2}\llik)^{-1}(\map)\|_{op} +  \|(\grad^{2}\llik)^{-1}(\param) - (\grad^{2}\llik)^{-1}(\map)\|_{op}  \\
&\le  \|(\grad^{2}\llik)^{-1}(\map)\|_{op} +  \|\grad (\grad^{2}\llik)^{-1}(\param)\|_{op} \|\param - \map\|_{2} \\
&\le  \frac{cd}{N} +  \frac{c^{2}L}{6\sqrt{3} N}  \|\param - \map\|_{2} ,
\]
so $\llik(\param)$ is $\alpha N/(cd)$-strongly concave for all $\param \in \ball_{\Delta}(\map)$ if 
\[
\frac{cd}{N} +  \frac{c^{2}L\Delta}{6\sqrt{3} N} \le \frac{cd}{N\alpha} 
\quad \iff \quad
\Delta^{2} \le \frac{108 d^{2}(1-\alpha)^{2}}{L^{2}c^{2}\alpha^{2}}.
\]
To apply \cref{thm:approx-map}, we require that $\Delta^{2} \ge 4\veps c d /\alpha$.
Combining the two inequalities, we have 
\[
\frac{4\veps c d}{\alpha} \le \frac{108 d^{2}(1-\alpha)^{2}}{L^{2}c^{2}\alpha^{2}}
\quad \iff \quad 
\frac{\veps c^{3}L^{2}}{27d}\alpha \le (1-\alpha)^{2}
\quad \iff \quad  
0 \le \alpha^{2} - (2 + b)\alpha + 1.
\label{eq:alpha-ineq}
\]
Solving the quadratic implies that the maximal viable $\alpha$ value is 
$\alpha^{*} = 1 + b - \sqrt{(b+1)^{2} - 1} \ge \frac{1}{2(b+1)}$. 

Requiring $R - \|\map\|_{2} \ge 2\sqrt{\frac{cd \veps}{\alpha^{*}}}$
together with the hypothesis that $\|\xn\| \le 1$ ensures that we are considering only inner products $\xn \cdot \param \in [-R, R]$. 
Since \cref{eq:bounding-assumption} holds by hypothesis, Assumption \ref{asm:allik-lb} holds.
The result now follows from \cref{thm:approx-map}. 
\eprf

\bprf[Proof sketch of \cref{cor:huber-map-error}]
The proof is similar in spirit to \cref{cor:logit-map-error-general}. 
The key differences are that we apply \cref{cor:smoothed-huber-chebyshev-error} and use the condition that
a constant fraction of the data satisfies $|\xn \cdot \map - \yn| \le b/2$ to guarantee $\Theta(N)$-strong log-convexity
of $-\log \post$ near the MAP. 
\eprf

Recall that a centered random variable $X$ is said to be \emph{$\sigma^{2}$-subgaussian}~\citep[Section 2.3]{Boucheron:2013} if for all $s \in \reals$,
\[
\EE[e^{sX}] \le e^{s^{2}\sigma^{2}/2}.
\]

\bnthm \label{thm:approx-post}
Assume that
\benum[resume*=assumptions]
\item \label{asm:normal-post} $-\log\apost(\param)$ is $\tilde\varrho$-strongly convex, %
\item \label{asm:xn-norm} for all $n=1,\dots,N$, $\|\xn\|_{2} \le 1$, 
\item \label{asm:error-growth} there exist constants $a_{n}, b, R, \alpha \in \reals_{+}$ such that 
\[
\|\grad_{\param}\glmMap(\ip{\yn\xn}{\param}) - \grad_{\param}\glmMapM(\ip{\yn\xn}{\param})\|_{2} \le a_{n} + b \max(0, |\ip{\yn\xn}{\param}| - R), \text{ and}
\]
\item \label{asm:post-strongly-convex} $-\log\post(\param)$ is $\varrho$-strongly convex with mean $\mean$.
\eenum
Let $\sigma_{1}, \sigma_{2}$ be the subgaussianity constants of, respectively, the random variables $\ip{\yn\xn}{\mean} - \delta_{1}$ 
and $\|\yn\xn\|_{2}^{2} - \delta_{2}$, where the randomness is over $n \dist \distUnif\theset{1,\dots,N}$.
Let $\delta_{1} \defined \EE[\ip{\yn\xn}{\mean}]$, $\delta_{2} \defined \EE[\|\yn\xn\|_{2}^{2}]$, and
$\bar a \defined \sum_{n=1}^{N} a_{n}$. 
Then there exists an explicit constant $\veps$ (equal to zero if $b=0$ and depending on $R$, $\varrho$, $\sigma_{1}$, $\sigma_{2}$, $\delta_{1}$,
and $\delta_{2}$ otherwise) such that
\[
\dw(\post, \apost) \le \tilde\varrho^{-1}(\bar a + N b\veps).
\]
\enthm
\brmk[Value of $\veps$]
The definition of the constant $\veps$ is given in the proof of the theorem. 
\ermk
\brmk[Assumptions]
Our posterior approximation result primarily depends on the peakedness of the approximate posterior (Assumption \ref{asm:normal-post}) and the error of 
the approximate gradients (Assumption \ref{asm:error-growth}).
If the gradients are poorly approximated then the error can be large 
while if the (approximate) posterior is flat then even small likelihood errors could lead to large shifts in expected values of the parameters
and hence large Wasserstein error. 
\ermk
\brmk[Verifying assumptions]
In the corollaries we use \cref{thm:chebyshev-deriv-approx} to control the gradient error in the case of Chebyshev 
polynomial approximations, which allows us to satisfy Assumption \ref{asm:error-growth}.  
Whether Assumption \ref{asm:normal-post} holds will depend on the choices of $M$, $\glmMap$, and $\prior$. 
For example, if $M = 2$ and $-\log\prior$ is convex, then the assumption holds. 
This assumption could be relaxed to only assume, e.g., a ``bounded concavity'' condition along with strong convexity
in the tails. 
See \citet{Eberle:2015}, \citet[Section 4]{Gorham:2016b}, and \citet[Appendix A]{Huggins:2017} for full details. 
It is possible that Assumption \ref{asm:post-strongly-convex} could also be weakened. 
The key is to have some control of the tails of $\post$. 
Both  $\ip{\yn\xn}{\mean}$ and $\|\yn\xn\|_{2}^{2}$ are subgaussian since $\yn\xn$ is bounded. 
\ermk

\bprf[Proof of \cref{thm:approx-post}]
By Assumption \ref{asm:error-growth}, we have that 
\[
\err(\param) &\defined
\|\grad\log\post(\param) - \grad\log\apost(\param)\|_{2} \\
&\le \sum_{n=1}^{N}\|\grad_{\param}\glmMap(\ip{\yn\xn}{\param}) - \grad_{\param}\glmMapM(\ip{\yn\xn}{\param})\|_{2} \\
&\le \bar a + \sum_{n=1}^{N}b\max(0, |\ip{\yn\xn}{\param}| - R).
\]
By \cref{lem:ip-subexponential}, the random variable $W \defined \ip{\yn\xn}{\param} - \delta_{1}$ is $(\lambda, \beta)$-subexponential.
Hence for $t \ge 0$, 
\[
\Pr(W \ge t) \maxop \Pr(W - \delta \le -t) 
\le \bar p(t, \lambda, \beta)
\defined e^{-\left(\frac{t^{2}}{2\lambda^{2}} \minop \frac{t}{2\beta}\right)}.
\]
We can now bound $\post(\err)$:
\[
\post(\err) 
&\le aN + \sum_{n=1}^{N}\EE_{\param \dist \post}[b\max(0, |\ip{\yn\xn}{\param}| - R)]. \\
&= aN + bN\EE_{n \dist \distUnif\theset{1,\dots,N}}\EE_{\param \dist \post}[\max(0, |\ip{\yn\xn}{\param}| - R)) \\
&= aN + bN\EE[\max(0, |W + \delta_{1}| - R))] \\
&= aN + bN\EE[(W+\delta_{1}+R)\ind(W + \delta_{1} \le -R) + (W + \delta_{1}-R)\ind(W + \delta_{1} \ge R)]. \label{eq:pi-err-intermediate-expectation}
\]
For the second term in the expectation, we have
\[
\lefteqn{\EE[(W + \delta_{1} - R)\ind(W \ge R - \delta_{1})]} \\
&= \int_{R - \delta_{1}}^{\infty}(w + \delta_{1} - R)p(\dee w) \\
&= \int_{R - \delta}^{\infty} \Pr(W \ge t)\,\dee t \\
&\le 0 \maxop (\delta_{1} - R) + \int_{0 \maxop (R - \delta_{1})}^{\infty} \bar p(t, \lambda, \beta) \dee t  =: B(R, \delta_{1},\lambda, \beta),
\]
By symmetry, the first term in the expectation in \cref{eq:pi-err-intermediate-expectation} is 
bounded by $B(R, -\delta_{1},\lambda, \beta)$, so 
\[
\post(\err) 
&\le \bar a + Nb(B(R, \delta_{1},\lambda, \beta) + B(R, -\delta_{1},\lambda, \beta)). 
\]

Assumption \ref{asm:normal-post} implies that $\apost$ satisfies Assumption 2.A of \citet{Huggins:2017} with $C = 1$ and $\rho = e^{-\tilde\varrho}$.
By Theorem 2 of \citet{Gorham:2016b}, it is not necessary for the Lipschitz conditions in Assumption 2.A of \citet{Huggins:2017} to hold. 
Furthermore, it can easily be seen that 2.B(3) of \citet{Huggins:2017} is not necessary if both $\post$ and $\apost$ are strongly convex. 
The remaining portions of Assumption 2.B of \citet{Huggins:2017} are satisfied, however. 
Thus we can apply Theorem 3.4 from \citet{Huggins:2017}, which yields
\[
\dw(\post, \apost) 
&\le \tilde\varrho^{-1}\post(\err) 
\le  \tilde\varrho^{-1}(\bar a + Nb \veps),
\]
where $\veps \defined B(R, \delta_{1},\lambda, \beta) + B(R, -\delta_{1},\lambda, \beta)$.
\eprf

\bnlem \label{lem:ip-subexponential}
Under the conditions of \cref{thm:approx-post}, the random variable $\ip{\yn\xn}{\param} - \delta_{1}$ is $(\lambda, \beta)$-subexponential, 
where $\lambda^{2} \defined 4 \left(\frac{1 + \delta_{2}}{\varrho} \maxop \sigma_{1}^{2}\right)$ and 
$\beta^{2} \defined \frac{2\sigma_{2}^{2}}{\varrho}$.
\enlem
\bprf
Let $\zn = \yn\xn$. 
For $|s| \le 1/\beta$, we have 
\[
\EE[e^{s(\ip{\zn}{\param} - \delta_{1})}]
&= \EE[\EE[e^{s\ip{\zn[]}{\param - \mean}} \given \zn = \zn[]] e^{s(\ip{\mean}{\zn} - \delta_{1})}] \\
&\le \EE[e^{s^{2}\|\zn\|_{2}^{2}/\varrho'} e^{s(\ip{\mean}{\zn} - \delta_{1})}]  & \text{Assumption \ref{asm:post-strongly-convex}} \\
&\le 0.5\EE[e^{2s^{2}\|\zn\|_{2}^{2}/\varrho'} + e^{2s(\ip{\mean}{\zn} - \delta_{1})}]  & \text{AM-GM inequality} \\
&\le 0.5[e^{4s^{4}\sigma_{2}^{2}/\varrho^{2}+2s^{2}\delta_{2}/\varrho} + e^{2s^{2}\sigma_{1}^{2}}] & \text{subgaussianity} \\
&\le 0.5[e^{2s^{2}(1 + \delta_{2})/\varrho} + e^{2s^{2}\sigma_{1}^{2}}]  & \text{bound on $|s|$} \\
&\le e^{s^{2}\lambda^{2}/2}.
\]
\eprf

\bncor \label{cor:approx-post-logistic}
Let $\glmMapM[2]$ be the second-order Chebyshev approximation to $\logitMap$ on $[-R,R]$ 
and let $\apost(\param) = \distNorm(\param \given \amap, \acov)$ denote the posterior approximation obtained by using $\glmMapM[2]$ 
with a Gaussian prior $\prior(\param) = \distNorm(\param \given \param_{0}, \bSigma_{0})$. 
Let $\mean \defined \int \param \post(\dee\param)$, let $\delta_{1} \defined N^{-1}\sum_{n=1}^{N}\ip{\yn\xn}{\mean}$,
and let $\sigma_{1}$ be the subgaussanity constant of the random variable $\ip{\yn\xn}{\mean} - \delta_{1}$, where $n \dist \distUnif\theset{1,\dots,N}$. 
Assume that $|\delta_{1}| \le R$, that $\|\acov\|_{2} \le cd/N$, and 
that $\|\xn\|_{2} \le 1$  for all $n=1,\dots,N$. 
Then with $\sigma_{0}^{2} \defined \|\bSigma_{0}\|_{2}$, we have 
\[
\dw(\post, \apost) \le cd\left(a(R) +  \sqrt{2}\sigma_{0} e^{8(2+\sigma_{1}^{2}\sigma_{0}^{-2})-\sqrt{2}\frac{R - |\delta_{1}|}{\sigma_{0}}}\right),
\]
where $a(R)$ is bounded by
\[
\min_{r \in (1,\pi/R  + \sqrt{\pi^{2}/R^{2} + 1})}\left|\log\left(1+ e^{-\frac{1}{2}R (r-r^{-1})\ii} \right)\right|\frac{(r+1)(9r^{2} + 7r + 2)}{r^{2}(r-1)^{4}}. \label{eq:logit-a-bound}
\]
\encor
\bprf %
Assumption \ref{asm:normal-post} holds by construction. 
The bound on 
\[
a(R) \defined \sup_{s \in [-R,R]}|\logitMap'(s) - \glmMapM[2]'(s)|
\]
follows immediately from \cref{cor:logit-chebyshev-error} in the case of $M=2$. 
Furthermore, since $\glmMapM[2]'(s) = b_{1,1} + b_{1,2}s$, for $|s| > R$, the additional error
is at most $|b_{1,2}|(|s|-R)$. 
In the case of a Chebyshev approximation, it is easy to verify that $|b_{1,2}| \le 0.25$ for all 
$R$ (since as $R \to 0$, $b_{1,2} \to \logitMap''(0) = -0.25$ and $-b_{1,2}$ is a decreasing function of $R$). 
In short, $|\logitMap'(s) - \glmMapM[2]'(s)| \le a(R) + 0.25\max(0, |s|-R)$ and therefore, 
using Assumption \ref{asm:xn-norm},  we have
\[
\lefteqn{\|\grad_{\param}\glmMap(\ip{\yn\xn}{\param}) - \grad_{\param}\glmMapM(\ip{\yn\xn}{\param})\|_{2}} \\
&= \|\glmMap'(\ip{\yn\xn}{\param})\yn\xn - \glmMapM'(\ip{\yn\xn}{\param})\yn\xn\|_{2} \\
&\le a(R) + .25\max(0, |\ip{\yn\xn}{\param}|-R).
\]
Hence Assumption \ref{asm:error-growth} holds with $a_{n} = a(R)$ and $b = 0.25$. 

Now, clearly $-\log\post$ is $\sigma_{0}^{-2}$-strongly convex.
Since  $\|\xn\|_{2} \le 1$, conclude that $\delta_{2} \le 1$ and $\sigma_{2} \le 1/2$. 
To upper bound $\veps$, note that
\[
B(R, \delta_{1},\lambda, \beta) + B(R, -\delta_{1},\lambda, \beta)
\le 2 B(R, |\delta_{1}|,\lambda, \beta)
\]
and that $\bar p(t, \lambda, \beta) \le e^{\frac{\lambda^{2}}{4\beta^{2}}}e^{-t/\beta}$. 
Also, $\lambda^{2} \le 4(2\sigma_{0}^{2} + \sigma_{1}^{2})$ and 
$\beta^{2} = \sigma^{2}_{0}/2$. 
Using this upper bound in $B(R, |\delta_{1}|,\lambda, \beta)$ along with straightforward 
simplifications yields:
\[
2B(R, |\delta_{1}a|,\lambda, \beta)
&\le 2\beta e^{\frac{\lambda^{2}}{4\beta^{2}}}e^{-\frac{R - |\delta_{1}|}{\beta}}
\le \sqrt{2}\sigma_{0} e^{8(2+\sigma_{1}^{2}\sigma_{0}^{-2})}e^{-\sqrt{2}\frac{R - |\delta_{1}|}{\sigma_{0}}}.
\]
The result now follows from \cref{thm:approx-post} since $-\log\apost$ is $\|\acov\|_{2}^{-1}$-strongly 
convex and hence by assumption $N/(cd)$-strongly convex. 
\eprf

\bncor \label{cor:approx-post-poisson}
Let $f_{M}(s)$ be the order-$M$ Chebyshev approximation to $e^{t}$ on the interval~${[-R,R]}$,
and let~$\apost(\param)$ denote the posterior approximation obtained by using the approximation
${\log \tp(\yn \given \xn, \param)  \defined  \yn \xn \cdot \param - f_{M}(\xn \cdot \param) - \log \yn!}$
with a log-concave prior on~${\paramspace = \ball_{R}(\bzero)}$.
If ${\inf_{s \in [-R,R]} f_{M}''(s) \ge \tilde\varrho > 0}$ and $\|\xn\|_{2} \le 1$  for all $n=1,\dots,N$, then
with~${\tau \defined \|\sum_{n=1}^{N} \xn\xn^{\top}\|_{2}}$, we have
\[
\dw(\post, \apost) \le \frac{N}{\tilde\varrho \tau} \min_{r>1}e^{\frac{1}{2}R (r+r^{-1})} \frac{(r+1)[M^{2}r(r+1) + M(2r^{2} + r + 1) + r(r+1)]}{r^{M}(r-1)^{4}}.
\] 
\encor
Note that ${\inf_{s \in [-R,R]} f_{M}''(s) \ge \tilde\varrho > 0}$ holds as long as $M$ is even and sufficiently large. 
\bprf %
Since by hypothesis ${\inf_{s \in [-R,R]} f_{M}''(s) \ge \tilde\varrho > 0}$, the prior is log-concave, 
and $-\log \apost$ is $\tilde\varrho \tau$-strongly convex (i.e., Assumption \ref{asm:normal-post} holds).
Using Assumption \ref{asm:xn-norm}, we have
\[
\lefteqn{\|\grad_{\param}\log p(\yn \given \xn, \param) - \grad_{\param}\log \tp(\yn \given \xn, \param) \|_{2}} \\
&= \|e^{\ip{\yn\xn}{\param})}\yn\xn - f_{M}'(\ip{\yn\xn}{\param})\yn\xn\|_{2} \\
&\le \sup_{s \in [-R,R]}|e^{-s}- f_{M}'(s)| =: a(R).
\]
which is bounded according to \cref{cor:exp-chebyshev-error}. 
Hence Assumption \ref{asm:error-growth} holds with $a_{n} = a(R)$ and $b = 0$.
The result now follows immediately from \cref{thm:approx-post}. 
\eprf

}

\bibliographystyle{abbrvnat}
{\small
\bibliography{library,refs}}

\end{document}